\documentclass[a4paper,11pt]{article}
\usepackage{}

\usepackage{jheppub} % for details on the use of the package, please
\usepackage[T1]{fontenc} % if needed
\usepackage{appendix}
\usepackage{mathtools}%
\usepackage{mathrsfs}%
\usepackage{amsfonts,amssymb}%гд
\usepackage{amsmath}%American Mathematical Society
\usepackage{tabularx}%
\usepackage{slashed}
\usepackage{cases}
\usepackage{bm}%гд
\usepackage{amsxtra}
\usepackage[dvipsnames]{xcolor}
\usepackage{colordvi}
\usepackage{colortbl}
\usepackage{float}
\usepackage{floatflt}
\usepackage{graphicx}
\usepackage{hhline}
\usepackage{mathrsfs}

\graphicspath{{figures/}}

\newcommand{\ben}{\begin{eqnarray}}
\newcommand{\een}{\end{eqnarray}}

\newcommand{\bef}{\begin{figure}[!htp]}
\newcommand{\eef}{\end{figure}}

\newcommand{\bea}{\begin{eqnarray}}
\newcommand{\eea}{\end{eqnarray}}

\def\ba{\begin{linenomath*}\begin{equation}}
\def\ea{\end{equation}\end{linenomath*}}

\allowdisplaybreaks

\newcommand{\state}[4]{{^{#1}\hspace{-0.6mm}#2_{#3}^{[#4]}}}
\newcommand{\stateprime}[4]{{^{#1}\hspace{-0.6mm}#2_{#3}^{\prime[#4]}}}

\newcommand\CScSa{\state{3}{S}{1}{1}}

\newcommand\COaSz{\state{1}{S}{0}{8}}

\newcommand\COcSa{\state{3}{S}{1}{8}}
\newcommand\COcPz{\state{3}{P}{0}{8}}

\newcommand\COcPj{\state{3}{P}{J}{8}}

\newcommand\lrd{\overleftrightarrow{D}}
\newcommand\lrde{\overleftrightarrow{\boldsymbol{\nabla}}}

\title{\boldmath \centering Color-octet contributions for $J/\psi$ inclusive production at B factories in soft gluon factorization}

\author[a]{An-Ping Chen}
\author[b,c]{Xiao-Bo Jin}
\author[b,d,e]{Yan-Qing Ma}
\author[b]{Ce Meng}
\affiliation[a]{College of Physics and Communication Electronics, Jiangxi Normal University, Nanchang 330022, China}
\affiliation[b]{School of Physics and State Key Laboratory of Nuclear Physics and
	Technology, Peking University, Beijing 100871, China}
\affiliation[c]{Center of Advanced Quantum Studies, Department of Physics,Beijing Normal University, Beijing 100875, China}
\affiliation[d]{Center for High Energy physics, Peking University, Beijing 100871, China}
\affiliation[e]{Collaborative Innovation Center of Quantum Matter,
	Beijing 100871, China}
\emailAdd{chenanping@jxnu.edu.cn}
\emailAdd{xiaobojin@pku.edu.cn}
\emailAdd{yqma@pku.edu.cn}
\emailAdd{mengce75@pku.edu.cn}

\abstract{We have studied color-octet contributions for $J/\psi$ inclusive production at B factories, i.e., $e^+e^-\to J/\psi(\COcPj,\COaSz) + X_{\mathrm{non}-c\bar c}$, using the soft gluon factorization (SGF) approach, in which the $J/\psi$ energy spectrum is expressed in a form of perturbatively calculable short-distance hard parts convoluted with one-dimensional soft gluon distributions (SGDs). The series of velocity corrections originated from kinematic effect can be naturally resummed in this approach.  Short-distance hard parts have been calculated analytically to next-to-leading order in $\alpha_s$.  Renormalization group equations for SGDs have been derived and solved, which resums  Sudakov logarithms originated from soft gluon emissions. Our final result gives a upper bound for color-octet matrix elements consistent with that extracted from hadron colliders. This may relieve the well-known universality problem in the NRQCD factorization.
	
As a comparison, we also analytically calculated short-distance hard parts in the NRQCD factorization, with Sudakov logarithms resummed by using soft collinear effective theory. The comparison shows that velocity corrections from kinematic effect, which have been resummed in SGF, are significant for phenomenological study. Furthermore, it is found that Sudakov logarithms originated from soft gluon emissions are very important, while it is not the case for Sudakov logarithms originated from jet function. Therefore, the partial Sudakov resummation in SGF has already captured the main physics.
 }


\begin{document}
\maketitle

\section{Introduction}

Although heavy quarkonium production has been widely studied in the nonrelativistic quantum chromodynamics (NRQCD) factorization~\cite{Bodwin:1994jh}, the underline mechanism is still under debate. The reason is that the NRQCD factorization can not provide a universal description of all quarkonium production data. In other words,  long-distance matrix elements (LDMEs) in NRQCD are found to be not universal.
It was argued in ref.~\cite{Ma:2017xno} that the universality problem in NRQCD may be caused by the bad convergence of velocity expansion, which suffers from large high
order relativistic corrections due to soft hadrons emission in the hadronization process. Resummation of these relativistic-correction terms will result in the so called soft gluon factorization (SGF) framework~\cite{Ma:2017xno}. It was demonstrated in ref.~\cite{Chen:2020yeg} that the SGF is equivalent to the NRQCD factorization, but with a series of important relativistic corrections originated from kinematic effects resummed. As a result, the SGF approach should has a much better convergence in the velocity expansion, and thus may provide a reasonable description of heavy quarkonium production.

Besides exclusive processes \cite{Li:2019ncs}, the SGF approach has been recently applied to calculate the fragmentation function of the gluon to a $\COcSa$ heavy
quark-antiquark pair in ref. \cite{Chen:2021hzo}, which was expressed in a form of perturbative short-distance hard part convoluted
with one-dimensional $\COcSa$ soft gluon distribution (SGD). With a NLO calculation of the short-distance hard part, the authors demonstrated that the SGF is valid at NLO level. The renormalization group equation of the $\COcSa$ SGD was derived and solved, which resummed Sudakov logarithms to all orders in perturbation theory. A comparison with gluon fragmentation function calculated in NRQCD factorization indicates that the
SGF formula resums a series of velocity corrections in NRQCD which are important for
phenomenological study.

Color-octet (CO) contributions of the $J/\psi$ production in $e^+e^-$ annihilation, i.e., $e^+e^-\to J/\psi(\COcPj,\COaSz) + X_{\mathrm{non}-c\bar c}$, play an important role to understand the production mechanism of quarkonium. The $J/\psi$ inclusive production at B factories have been measured by the BaBar and Belle collaborations \cite{BaBar:2001lfi,Belle:2001lqi,Belle:2002tfa,Belle:2009bxr} and have been studied in NRQCD factorization extensively \cite{Kiselev:1994pu,Braaten:1995ez,Yuan:1996ep,Cho:1996cg,Baek:1998yf,Schuler:1998az,Liu:2003zr,Liu:2003jj,Zhang:2006ay,Gong:2009ng,Ma:2008gq,Gong:2009kp,He:2009uf,Jia:2009np,Zhang:2009ym}.
In NRQCD factorization, the cross section $\sigma[e^+e^-\to J/\psi + X_{\mathrm{non}-c\bar c} ]$ at leading order (LO) in $\alpha_s$ includes color-singlet (CS) contribution $e^+e^-\to J/\psi(\CScSa) + gg $, and
CO contributions $e^+e^-\to J/\psi(\COcPj,\COaSz) + g$. For $J/\psi$ energy spectrum, the LO CO contribution predicts an apparent enhancement at the $J/\psi$ maximum energy \cite{Braaten:1995ez}, but experiments did not show any enhancement at the endpoint region. It was then clear that large Sudakov logarithms appear at higher order invalidate the perturbative expansion of CO contributions. The behavior at the endpoint region can be qualitatively explained if the resummation of the Sudakov logarithms as well as nonperturbative effects are considered \cite{Beneke:1997qw,Fleming:2003gt}. On the other hand, the Belle measurement gives
\begin{align}
\sigma[e^+e^-\to J/\psi + X_{\mathrm{non}-c\bar c} ]=(0.43\pm0.09\pm0.09) \text{ pb},
\end{align}
which can be well saturated by CS channel with the next-to-leading order (NLO) in $\alpha_s$ correction \cite{Ma:2008gq,Gong:2009kp}, $\mathcal {O}(v^2)$ relativistic correction \cite{He:2009uf,Jia:2009np} and QED initial-state radiation effect \cite{Shao:2014rwa},
leaving little
room for the contribution of CO channel.
Even though, an upper bound of CO LDMEs can be obtained by setting the CS contribution to be zero, which at NLO level gives \cite{Zhang:2009ym}
\begin{align}
\langle \mathcal{O}^{J/\psi}(\COaSz) \rangle + 4.0\frac{\langle \mathcal{O}^{J/\psi}(\COcPz) \rangle}{m_c^2} <(2.0\pm0.6)\times 10^{-2}\text{GeV}^3.
\end{align}
However, this upper bound is much smaller than the value of CO LDMEs extracted from  hadron colliders \cite{Ma:2010yw,Butenschoen:2010rq,Gong:2012ug,Bodwin:2014gia,Faccioli:2014cqa}, which challenges the universality of LDMEs.

Because SGF has a better convergence in velocity expansion, in this paper we apply it to study  the CO contributions of $e^+e^-\to J/\psi(\COcPj,\COaSz) + X_{\mathrm{non}-c\bar c}$.
The rest of the paper is organized as follows. In section \ref{sec:SGF}, we give a short review of SGF formula. Especially, we introduce a new lower cutoff $x_{\textrm{min}}$ of momentum fraction and demonstrate that final result is insensitive to the value of $x_{\textrm{min}}$.
In section \ref{sec:hard-part}, we present perturbative calculation in SGF. We also  discuss the renormalization group equation (RGE) of SGDs.
In section \ref{sec:NRQCD}, we present perturbative calculation in NRQCD, and use the soft collinear effective theory (SCET)~\cite{Bauer:2000ew,Bauer:2000yr,Bauer:2001ct,Bauer:2001yt,Fleming:2003gt} to resum large Sudakov logarithms to the NLL accuracy.

A comparison of phenomenological results obtained in SGF and NRQCD factorization are presented in section \ref{sec:Phenomenology} and a summary is given in section \ref{sec:summary}. In appendix~\ref{ap:expression} we list some analytical expressions in perturbative calculation. Finally, we discuss the $x_{\textrm{min}}$ dependence in appendix~\ref{ap:cutoff-dependence}.

\section{Soft gluon factorization formula}\label{sec:SGF}

We denote a four-vector $a$ as
 \begin{align}
  a^\mu= (a^0,a^1,a^2,a^3)=(a^0,\boldsymbol{a}).\nonumber
\end{align}
 We also use light-cone coordinates where a four-vector $a$ can be expressed as
\begin{align}
a^\mu&=(a^+,a^-,a^1,a^2)= (a^+,a^-,a_\perp),\nonumber\\
a^+ &= (a^0+a^3)/\sqrt{2},\nonumber\\
a^- &= (a^0-a^3)/\sqrt{2},\nonumber
\end{align}
where the subscript $\perp$ denotes the perpendicular direction. Then scalar product of two four-vector $a$ and $b$ becomes
\begin{align}
a \cdot b = a^+ b^- + a^- b^+ + a_\perp \cdot b_\perp. \nonumber
\end{align}
We introduce a light-like vector $l^\mu = (0, 1, 0_\perp)$, so that $a\cdot l=a^+$.

Then in the rest frame of $J/\psi$ we have
\begin{equation}
P_\psi^\mu=(M_\psi,\boldsymbol{0}),\quad q^\mu=(0,\boldsymbol{q}), \nonumber
\end{equation}
where $P_\psi$ is the momentum of $J/\psi$, $M_\psi$ is the mass of $J/\psi$ and $q$ is the  half of the relative momentum of the heavy quark-antiquark pair in $J/\psi$, which is relate to the relative velocity $v$ by $v=|\boldsymbol{q}|/m_c$, where $m_c$ is the charm quark mass.

The SGF is equivalent to the NRQCD factorization but with a series of important relativistic corrections originated from kinematic effects resummed~\cite{Chen:2020yeg}. Beginning from the leading operator for a specific quantum number in NRQCD Lagrangian, e.g., $\psi^\dag \chi$ (or  $\psi^\dag \sigma^i T^a \chi$, $\psi^\dag \lrde^i \chi$, and so on), one can construct powers suppressed operators like $\psi^\dag \lrde^2 \chi$, $\boldsymbol{\nabla}^2(\psi^\dag \chi)$ or $\psi^\dag g \boldsymbol{E}\cdot \sigma \chi$ by inserting the relative derivative $\lrde^2$, the total derivative $\boldsymbol{\nabla}^2$, or the $\boldsymbol{E}$ and $\boldsymbol{B}$ fields. The first two kinds of insertions are originated from kinematic effects and  they are chosen to be resummed. Using equations of motion, one can replace the relative derivatives $\lrde^2$ and $\lrde_0$ by total derivatives, which results in operators like $\boldsymbol{\nabla}_0^{n_1}\boldsymbol{\nabla}^{2n_2}(\psi^\dag  \chi)$. Then using integration by parts, one can eliminate all operators except that with $n_1=n_2=0$. The price to pay is the introduction of the relative momentum between physical quarkonium and the intermediate heavy quark-antiquark pair into short-distance hard parts~\cite{Chen:2020yeg}. The final formula
is the SGF proposed in ref.~\cite{Ma:2017xno}. Note that, in the above derivation, one needs to introduce proper gluon fields to combine with spacetime derivatives to form gauge covariant derivatives.

In SGF, the production cross section of $J/\psi$ in $e^+e^-$ collisions can be expressed in following formula~\cite{Ma:2017xno}:
\begin{align}\label{eq:fac1d-1}
 (2\pi)^3 2 P_{\psi}^0 \frac{\mathrm{d}\sigma_{\psi}}{\mathrm{d}^3P_{\psi}}
 &= \sum_{n,n^\prime} \int_{x_{\mathrm{min}}}^1 \frac{\mathrm{d}x}{x^2}   \hat \sigma_{[n n^\prime ]}(P_{\psi}/x, s, m_c, x_{\mathrm{min}}/x, \mu_f)    F_{[n n^\prime] \rightarrow \psi }(x,M_{\psi},m_c, \mu_f),
\end{align}
where $x$ is the longitudinal momentum fraction defined as $x = P^+_\psi/P^+$ with $P_{\psi}$ denoting the momentum of $J/\psi$ and $P$ denoting the total momentum of the intermediate $c\bar c$ pair. Different from the original form, we have introduced a lower cutoff $x_{\mathrm{min}}$ for the $x$ integration. This is allowed because, if $x$ is too small, the intermediate $c\bar c$ pair will emit hard gluons during the hadronization process, which effect is perturbatively calculable. In appendix \ref{ap:cutoff-dependence} we demonstrate that the production cross section is insensitive to the value of $x_{\mathrm{min}}$. In other words, the $x_{\mathrm{min}}$ dependence of the integration limit cancels with the $x_{\mathrm{min}}$ dependence of  $\hat{\sigma}_{[n n^\prime]}$. This is not surprised at all because  $\hat{\sigma}_{[n n^\prime]}$ is determined by matching the both sides of the above factorization formula. Therefore, as a special choice, one can also set $x_{\mathrm{min}}=0$~\cite{Ma:2017xno}.

In eq. \eqref{eq:fac1d-1}, $\mu_f$ is the factorization scale, $\sqrt{s}$ is the center-of-mass energy of the $e^+e^-$ system, and the quarkonium state created by $a_\psi^\dag$ has
standard relativistic normalization. $\hat{\sigma}_{[n n^\prime]}$ are perturbatively calculable short-distance hard parts that produce an intermediate $c \bar c$ pair with quantum numbers $n=\state{{2S+1}}{L}{J,J_z}{c}$ and $n^\prime =\stateprime{{2S^\prime+1}}{L}{J^\prime,J_z^\prime}{c^\prime}$ in the amplitude and the complex-conjugate of the amplitude, respectively, with $c,c^\prime = 1 $ or $8$ representing the color-singlet or color-octet state of the $c \bar c$ pair. $F_{[n n^\prime] \to \psi }$ are one-dimensional SGDs which are defined explicitly as~\cite{Chen:2021hzo}
\begin{align}\label{eq:SGD1d-1}
F_{[n n^\prime] \to \psi}(x,M_\psi,m_c, \mu_f)
&= P_\psi^+\int \frac{\mathrm{d}b^-}{2\pi} e^{-iP_\psi^+  b^-/x} \langle 0| [\bar\Psi \mathcal {K}_{n} \Psi]^\dag(0) [a_\psi^\dag a_\psi] [\bar\Psi \mathcal {K}_{n^\prime}\Psi](b^-) |0\rangle_{\textrm{S}},
\end{align}
where $\Psi$	denotes the Dirac field of heavy quark. The subscript ``S'' means that the field operators in the definition are the operators obtained in small momentum region. In
additional, we define ``S'' to select only leading power terms in threshold expansion\cite{Chen:2021hzo}, that is the expansion in the limit
$(P-P_\psi)^+=(1-x)P^+\to0$.

In general the state $n$ can be different from the state $n^\prime$, but for the
case of producing a polarization summed $J/\psi$, there are
constraints $c=c^\prime, S=S^\prime, J =J^\prime, J_z=J_z^\prime$ and $|L-L^\prime|=0,2,4,\cdots$~\cite{Ma:2017xno,Ma:2015yka}. While in this work, we only consider the case $n=n^\prime$ with $n=\state{{1}}{S}{0}{8}$ or $ \state{{3}}{P}{J,\lambda}{8}$, where are the most important color-octet contributions for $J/\psi$ production at $e^+e^-$ collision. The corresponding   projection operators $\mathcal {K}_{n}$, which define the intermediate state $n$, are given by~\cite{Ma:2017xno}
\begin{subequations}\label{eq:projection1}
\begin{align}
 \mathcal {K}_{\state{{1}}{S}{0}{8}}(rb^-) =& \frac{\sqrt{M_ {\psi}}}{M_ {\psi}+2 m_c}\frac{M_ {\psi} + \slashed{P}_ {\psi}}{2M_ {\psi}} \mathcal {C}^{[8]}_{ a} \gamma_5 \frac{M_ {\psi} - \slashed{P}_ {\psi}}{2M_ {\psi}} ,
 \\
 \mathcal {K}_{\state{{3}}{P}{J,\lambda}{8}}(rb^-) =& \frac{\sqrt{M_ {\psi}}}{M_ {\psi}+2 m_c}\frac{M_ {\psi} + \slashed{P}_ {\psi}}{2M_ {\psi}} \mathcal {C}^{[8]}_{ a} \mathcal {E}_{J,\lambda}^{\mu\nu} \gamma_\mu \biggr(-\frac{i}{2}\biggr) \lrd_\nu \frac{M_ {\psi} - \slashed{P}_ {\psi}}{2M_ {\psi}},
\end{align}
\end{subequations}
where $D_\mu$ is the gauge covariant derivative with $\overline\Psi \lrd_\mu \Psi =
\overline\Psi (D_\mu \Psi) -
(D_\mu \overline\Psi)\Psi$. The color operator is given by
\begin{align}
\mathcal {C}^{[8]}_{ a}=\sqrt{2}T^{ \bar a} \Phi_{l}(rb^-)_{\bar a  a},
\end{align}
 with gauge link $\Phi_{l}(rb^-)_{\bar a   a}$ defined along the $l^\mu$ direction,
\begin{equation}\label{eq:gaugelink}
  \Phi_l(rb^{-})= \mathcal {P} \, \text{exp} \left[-i g_s
  \int_{0}^{\infty}\mathrm{d}\xi l\cdot A(rb^{-} + \xi l) \right] \, ,
\end{equation}
where $\mathcal {P}$ denotes path ordering and $A^{\mu}(x)$ is gluon field in the adjoint representation: $[A^{\mu}(x)]_{ac} = i f^{abc} A^{\mu}_{b}(x)$.  $\mathcal {E}_{J,\lambda}$ are  polarization tensors for $\state{{3}}{P}{J,\lambda}{8}$ states, with the following summation rules,
\begin{subequations}
\begin{align}
 \sum_{\lambda} \mathcal {E}_{0,\lambda}^{\mu\nu} \mathcal {E}_{0,\lambda}^{\ast\mu^\prime \nu^\prime} &=\frac{1}{d-1} \mathbb{P}^{\mu\nu}\mathbb{P}^{\mu^\prime \nu^\prime} , \\
\sum_{\lambda} \mathcal {E}_{1,\lambda}^{\mu\nu} \mathcal {E}_{1,\lambda}^{\ast\mu^\prime \nu^\prime} &=\frac{1}{2}[ \mathbb{P}^{\mu\mu^\prime}\mathbb{P}^{\nu \nu^\prime}-\mathbb{P}^{\mu\nu^\prime}\mathbb{P}^{\mu^\prime \nu}],\\
\sum_{\lambda} \mathcal {E}_{2,\lambda}^{\mu\nu} \mathcal {E}_{2,\lambda}^{\ast\mu^\prime \nu^\prime} &=\frac{1}{2}[ \mathbb{P}^{\mu\mu^\prime}\mathbb{P}^{\nu \nu^\prime}+\mathbb{P}^{\mu\nu^\prime}\mathbb{P}^{\mu^\prime \nu}]-\frac{1}{d-1} \mathbb{P}^{\mu\nu}\mathbb{P}^{\mu^\prime \nu^\prime},\\
\sum_{J,\lambda} \mathcal {E}_{J,\lambda}^{\mu\nu} \mathcal {E}_{J,\lambda}^{\ast\mu^\prime \nu^\prime} &= \mathbb{P}^{\mu\mu^\prime}\mathbb{P}^{\nu \nu^\prime},
\end{align}
\end{subequations}
where $d$ is the space-time dimension, the spin projection operator $\mathbb{P}^{\mu\nu}$ is defined as
\begin{align}
\mathbb{P}^{\mu\nu}=-g^{\mu\nu}+\frac{P_{\psi}^\mu P_{\psi}^\nu}{M_\psi^2}.
\end{align}

For energy distribution, we can rewrite eq.~\eqref{eq:fac1d-1} as
 \begin{align}\label{eq:SGF-form}
\frac{\mathrm{d}\sigma_{J/\psi}}{\mathrm{d}z}
=&  \sum_{n} \int_{\mathrm{max}[(z+\sqrt{z^2-4r})/2,x_{\mathrm{min}}]}^1 \frac{\mathrm{d}x}{x}   H_{[n]}(\hat z, M_\psi/x, s, m_c, x_{\mathrm{min}}/x, \mu_f) \nonumber\\
&\times F_{[n] \to \psi }(x,M_{\psi},m_c, \mu_f),
\end{align}
where we denote $[n] \equiv [nn]$, with $n=\state{{1}}{S}{0}{8}$ or $ \state{{3}}{P}{J,\lambda}{8}$, and new variables are defined as
\begin{align}
z\equiv \frac{2E_\psi}{\sqrt{s}}, \quad \hat z\equiv \frac{z}{x}= \frac{2E_\psi}{x\sqrt{s}},  \quad r\equiv \frac{M_\psi^2}{s}.
\end{align}
The variables $E_{\psi}/x$ and $M_\psi/x$ in the hard part correspond to energy $E_{c\bar c}$ and invariant mass  $M_{c\bar c}$ of the intermediate $c\bar c$ pair, respectively.

In above factorization formula, the short-distance hard parts $H_{[n]}$ are determined by the matching procedure  \cite{Ma:2017xno,Chen:2021hzo}. To this end, we replace the final-state $J/\psi$ by an on-shell $c\bar c$ pair with certain quantum number $n$ and momenta
\begin{align}
p_c= \frac{1}{2}P_\psi +q ,  \quad \quad p_{\bar c}= \frac{1}{2}P_\psi -q.
\end{align}
On-shell conditions $p_c^2= p_{\bar c}^2=m_c^2$ result in
\begin{align}
P_\psi \cdot q= 0 , \quad \quad q^2=m_c^2-P_\psi^2/4.
\end{align}
To project the final-state $c \bar c$ pair to the state $n$, we replace  spinors of the $c \bar c$ by following projector
\begin{align}\label{eq:spinor-operator}
\Pi[n]=\frac{2}{\sqrt{M_\psi}(M_\psi+2m_c)} ( \slashed{p}_{\bar c} - m_c )  \frac{M_\psi - \slashed{P}_\psi}{2M_\psi} \widetilde{\Gamma}^s_{n}
\widetilde{\mathcal {C}}^{[c]}
\frac{M_\psi + \slashed{P}_\psi}{2M_\psi} (\slashed{p}_{c} +m_c ),
\end{align}
where, for $n=\state{{1}}{S}{0}{8}$ or $\state{{3}}{P}{J,\lambda}{8}$, the color operator and spin operator are given by
\begin{subequations}
\begin{align}
\widetilde{\mathcal {C}}^{[8]}=&\sqrt{\frac{2}{N_c^2-1}}T^{ a}, \\
\widetilde{\Gamma}^s_{\state{{1}}{S}{0}{8}}=& \gamma_5, \\
\widetilde{\Gamma}^s_{\state{{3}}{P}{J,\lambda}{8}}=&  (d-1) \frac{q_\alpha }{\vert \textbf{q} \vert^2}  \mathcal {E}_{J,\lambda}^{\ast \alpha \mu}\gamma_\mu.
\end{align}
\end{subequations}
The factor $\sqrt{N_c^2-1}$ is to average over
color-octet states. We insert the perturbative expansions
\begin{subequations}
  \begin{align}\label{eq:diff order}
    \frac{\mathrm{d}\sigma_{c\bar c[n]}}{\mathrm{d}z}
    &=
    \,\frac{\mathrm{d}\sigma_{c\bar c[n]}^{LO}}{\mathrm{d}z}
    +
    \,\frac{\mathrm{d}\sigma_{c\bar c[n]}^{NLO}}{\mathrm{d}z} + \cdots,\\
    F_{[n^\prime] \rightarrow {c\bar c[n]}}
    &=
    F_{[n^\prime] \rightarrow {c\bar c[n]}}^{LO}
    +
    F_{[n^\prime] \rightarrow {c\bar c[n]}}^{NLO} + \cdots,\\
    H_{[n^\prime]}
    &=
    H_{[n^\prime]}^{LO}
    +
    H_{[n^\prime]}^{NLO} + \cdots
  \end{align}
\end{subequations}
into the factorization formula eq.~\eqref{eq:SGF-form}.
At leading order we have\cite{Ma:2017xno}
\begin{equation}\label{eq:SGD-LO0}
  F^{LO}_{[n^\prime]\rightarrow {c\bar c[n]}}(x,M_\psi,m_c,\mu_f) =\delta_{n^\prime n} \delta(1-x),
\end{equation}
which results in following matching relations up to NLO
\begin{subequations}\label{eq:matching-relation}
  \begin{align}
    H_{[n]}^{LO}(z,M_\psi,s,m_c,x_{\mathrm{min}},\mu_f)
    &=\frac{\mathrm{d}\sigma_{c\bar c[n]}^{LO}}{\mathrm{d}z} (z,M_\psi,s,m_c),
    \\
    H_{[n]}^{NLO}(z,M_\psi,s,m_c,x_{\mathrm{min}},\mu_f)
    &=
    \frac{\mathrm{d}\sigma_{c\bar c[n]}^{NLO}}{\mathrm{d}z} (z,M_\psi,s,m_c) \nonumber\\
    &\hspace{-4.5cm}
    - \sum_{n^\prime}\int_{\mathrm{max}[\frac{\sqrt{z^2-4r}+z}{2},
   x_{\mathrm{min}}]}^1
    \frac{\mathrm{d}x}{x}
    \frac{\mathrm{d}\sigma_{c\bar c[n^\prime]}^{LO}}{\mathrm{d}z} \left(\frac{z}{x},\frac{M_\psi}{x},s,m_c \right)
    F^{NLO}_{{[n^\prime]}\to {c\bar c[n]}}(x,M_\psi,m_c,\mu_f).
  \end{align}
\end{subequations}

In the SGF, velocity expansion is achieved by expanding $m_c^2$ in the hard parts $H_{[n]}$ around $M_\psi^2/4x^2$, which results in
 \begin{align}\label{eq:v-expansion}
 \frac{\mathrm{d}\sigma_{J/\psi}}{\mathrm{d}z}
 =& \sum_{i=0} \sum_{n} \int_{\mathrm{max}[(z+\sqrt{z^2-4r})/2,x_{\mathrm{min}}]}^1 \frac{\mathrm{d}x}{x}   H_{[n]}^{(i)}(\hat z,M_\psi/x,s,x_{\mathrm{min}}/x,\mu_f) \biggr( m_c^2 -\frac{M_\psi^2}{4x^2} \biggr)^i  \nonumber\\
 & \times F_{[n] \to \psi }(x,M_{\psi},m_c, \mu_f).
\end{align}
In ref.~\cite{Chen:2021hzo}, an explicit NLO calculation shows that the above kind of velocity expansion has good convergence, and the lowest order in the expansion can give a very good approximation of the full result. Therefore, to simplify the perturbative calculation, we only consider the contribution of $H_{[n]}^{(0)}$ here, with
 \begin{align}\label{eq:lowest-order1}
H_{[n]}^{(0)}(z/x,M_\psi/x,s,x_{\mathrm{min}}/x,\mu_f) =H_{[n]}(z/x,M_\psi/x,s,m_c, x_{\mathrm{min}}/x,\mu_f)\vert_{m_c=M_\psi/2x}.
\end{align}
Based on eqs.~\eqref{eq:lowest-order1} and~\eqref{eq:matching-relation}, we can calculate the short-distance hard parts perturbatively. Especially, we can set $m_c=M_\psi/2x$ in the integrand level before performing loop integration.

\section{Perturbative calculation of the short-distance hard parts}\label{sec:hard-part}

\subsection{The SGDs and the evolution equations }\label{sec:SGF resum}

To obtain short-distance hard parts $H_{[n]}^{(0)}$ up to NLO, we need first to calculate the perturbative SGDs.
One-loop correction for the perturbative SGDs can be derived by following the calculation details of color-octet $^3S_1$ SGD given in~\cite{Chen:2021hzo}. As we keep only the leading order in the velocity expansion, we can expand $m_c^2$ around $M_\psi^2/4$ before performing the loop integration and phase space integration. The calculation is straightforward and we get
\begin{subequations}\label{eq:SGDresult}
\begin{align}
F_{[\COaSz] \to c\bar c[\COaSz]}(x,M_\psi,m_c, \mu_f) =&  \delta(1-x) + \frac{\alpha_s N_c} {4\pi}
  \biggr[ \frac{4x}{(1-x)_+}  \ln\biggr(\frac{ x^2\mu_f^2 e^{ - 1}}{M_\psi^2}\biggr)
\nonumber\\
&\hspace{-4cm}
 - 8x
 \biggr( \frac{\ln(1-x)}{1-x} \biggr)_+
 - \delta(1-x)  \biggr( \ln^2\biggr(\frac{\mu_f^2 e^{ - 1}}{M_\psi^2}\biggr) +\frac{\pi^2}{6} -1 \biggr)
 \nonumber\\
&\hspace{-4cm}
- \frac{1}{N_c^2} \frac{\pi^2}{\Delta}\delta(1-x) \biggr] + \mathcal {O}(\Delta), \\
 F_{[\state{{3}}{P}{J^\prime,\lambda^\prime}{8}] \to c\bar c[\state{{3}}{P}{J,\lambda}{8}]}(x,M_\psi,m_c, \mu_f) =&  \delta_{J^\prime J}\delta_{\lambda^\prime \lambda}\biggr\{\delta(1-x) + \frac{\alpha_s N_c} {4\pi}
  \biggr[ \frac{4x}{(1-x)_+}  \ln\biggr(\frac{ x^2\mu_f^2 e^{ - 1}}{M_\psi^2}\biggr)
\nonumber\\
&\hspace{-4cm}
 - 8x
 \biggr( \frac{\ln(1-x)}{1-x} \biggr)_+
 - \delta(1-x)  \biggr( \ln^2\biggr(\frac{\mu_f^2 e^{ - 1}}{M_\psi^2}\biggr) +\frac{\pi^2}{6} -1 \biggr)
\nonumber\\
&\hspace{-4cm}
- \frac{1}{N_c^2} \frac{\pi^2}{\Delta}\delta(1-x) \biggr] \biggr\} + \mathcal {O}(\Delta).
\end{align}
\end{subequations}
Here $\Delta =\sqrt{1-4m_c^2/M_\psi^2}$, the ``plus'' functions are defined in the standard way through
\begin{align}\label{eq:plusfunction}
\int_z^1 \mathrm{d} x f(x)\biggr[\frac{\ln^m(1-x)}{1-x} \biggr]_+ = & \int_z^1 \mathrm{d} x [f(x) - f(1)]\frac{\ln^m(1-x)}{1-x} + \frac{f(1)}{m+1} \ln^{m+1}(1-z),
\end{align}
where $f(x)$ is a well-behaved regular function.

In SGF, the general form of RGEs for SGDs is given by~\cite{Chen:2021hzo}
\begin{align}\label{eq:general-RGE-SGDs}
\frac{\mathrm{d}}{\mathrm{d} \ln \mu_f}  F_{[n] \to \psi}(x,M_\psi,m_c, \mu_f)
=& \sum_{n^\prime} \int_x^1 \frac{\mathrm{d}y}{y} \boldsymbol{K}_{[n]}^{[n^\prime]}(x/y, M_\psi/y, m_c,\mu_f)
\nonumber\\
&\times F_{[n^\prime] \to \psi}(y,M_\psi,m_c, \mu_f),
\end{align}
where the evolution kernel can also be expressed in the velocity expansion series
 \begin{align}
 \boldsymbol{K}_{[n]}^{[n^\prime]}(x/y, M_\psi/y, m_c,\mu_f)
 =& \sum_{i=0} \boldsymbol{K}_{[n]}^{[n^\prime],(i)}( x/y,M_\psi/y,\mu_f) \biggr( m_c^2 -\frac{M_\psi^2}{4y^2} \biggr)^i.
\end{align}
Using above perturbative SGD results we can obtain evolution kernel $\boldsymbol{K}_{[n]}^{[n^\prime],(0)}$ at LO in $\alpha_s$ by matching both sides of the RGE,
\begin{subequations}
\begin{align}
\boldsymbol{K}_{[\COaSz]}^{[\COaSz],(0),LO}(x, M_\psi,\mu_f)=&\frac{\alpha_s }{4\pi} \biggr[ 2\Gamma_0^F \biggr ( \frac{ x}{(1- x)_+}  -\ln \frac{ \mu_f }{M_\psi}
\delta(1- x)\biggr ) +  \gamma^F_0 \delta(1- x) \biggr ], \\
\boldsymbol{K}_{[\state{{3}}{P}{J,\lambda}{8}]}^{[\state{{3}}{P}{J^\prime,\lambda^\prime}{8}],(0),LO}(x, M_\psi,\mu_f)=&\frac{\alpha_s }{4\pi} \biggr[ 2\Gamma_0^F \biggr ( \frac{ x}{(1- x)_+}  -\ln \frac{ \mu_f }{M_\psi}
\delta(1- x)\biggr ) + \gamma^F_0 \delta(1- x) \biggr ] \nonumber\\
&\times \delta_{JJ^\prime}\delta_{\lambda \lambda^\prime},
\end{align}
\end{subequations}
where
\begin{align}\label{eq:cusp}
\Gamma^F_0 = 4 N_c,  \quad \gamma^F_0 =& 4 N_c .
\end{align}

For the convenience of discussion, here we give some details of solving the above RGEs. We take $\COaSz$ SGD as an example. Following the discussion in ref. \cite{Chen:2021hzo}, we rewrite the RGE as a function of the variable
$\omega_0=(1-x)/x$, which is
\begin{align}\label{eq:RGE-SGD1}
& \frac{\mathrm{d}}{\mathrm{d} \ln\mu_f}
F_{[\COaSz] \to \psi}(\omega_0,M_\psi,m_c, \mu_f)
\nonumber\\
&= \int_0^{\omega_0} \mathrm{d} \omega^\prime_0  \frac{\alpha_s}{4\pi} \biggr [2\Gamma^F_0 \biggr ( \biggr[ \frac{1}{\omega_0-\omega^\prime_0} \biggr]_+ - \ln \frac{\mu_f}{M_\psi}
\delta(\omega_0-\omega^\prime_0)\biggr ) + \gamma^F_0 \delta(\omega_0 - \omega^\prime_0) \biggr]
\nonumber\\
& \quad \times
F_{[\COaSz] \to \psi}(\omega^\prime_0,M_\psi,m_c, \mu_f),
\end{align}
where $\omega^\prime_0=(1-y)/y$. $\omega_0$ (or $\omega^\prime_0$) is actually the longitudinal momentum fraction of the emitted soft gluons, and it takes values between $0$ and $\infty$. To solve the above equation, it is convenient to
perform the Laplace transformation, which, together with its inverse, is given by
\begin{equation}
\tilde{f}(\nu)= \int_0^\infty \mathrm{d}\omega_0 e^{- \omega_0 \nu}f(\omega_0) , \quad \textrm{and} \quad f(\omega_0)= \frac{1}{2\pi i}\int_{c-i \infty}^{c + i\infty} \mathrm{d} \nu e^{ \omega_0 \nu}\tilde{f}(\nu),
\end{equation}
where the constant $c$ is chosen to be larger than the real part of the rightmost singularity of $\tilde{f}(s)$. Performing a Laplace transform in eq.~\eqref{eq:RGE-SGD1} we obtain a RGE in Laplace space
\begin{align}\label{eq:LaplaceRGE}
\frac{\mathrm{d}}{\mathrm{d} \ln\mu_f}
\tilde F_{[\COaSz] \to \psi}(\nu , M_\psi,m_c, \mu_f)
=&  \frac{\alpha_s}{4\pi} \biggr(  -2\Gamma^F_{0}   \ln \frac{\bar \nu \mu_f}{M_\psi}
+ \gamma^F_0 \biggr) \nonumber\\
&  \times \tilde F_{[\COaSz] \to \psi}(\nu , M_\psi,m_c, \mu_f),
\end{align}
where $\bar \nu=\nu e^{\gamma_E}$, $\gamma_E$ is the Euler's constant. Solving the above equation, we then obtain \cite{Chen:2021hzo}
\begin{align}\label{eq:Resummed-SGD}
\tilde {F}_{[\COaSz] \to \psi}(\nu,M_\psi,m_c, \mu_r) =&     \biggr[1 + \frac{\alpha_s(\mu_r)N_c} {4\pi}
  \biggr(  -    4\ln^2\frac{\mu_r e^{\gamma_E}}{M_\psi} + 4\ln\frac{\mu_r e^{\gamma_E} }{M_\psi} - \frac{5\pi^2}{6}   \biggr)\biggr]
  \nonumber\\&\times
  \mathrm{exp}\biggr[ h_0(\mu_r,\chi_{0}) \biggr] \tilde F^{\mathrm{mod}}[\COaSz](\nu).
\end{align}
Here we evolved the SGD from the initial scale $\mu_r/\nu$ to the scale $\mu_r$. $\tilde F^{\mathrm{mod}}[\COaSz](\nu)$ is a model introduced to describe the nonperturbative effects at the initial scale $\mu_r/\nu$. And the evolution function $h_0$ is given by
\begin{align}\label{eq:sgfh}
h_0(\mu_r,\chi_{0})=&\frac{2\pi \Gamma^F_0}{\beta_0^2}   \frac{ 1}{\alpha_s(\mu_r)} \Big( -(1-2\chi_{0})\ln (1-2\chi_{0}) - 2\chi_{0} \Big)
- \frac{ \gamma_0^F}{2\beta_0} \ln (1-2\chi_{0}) \nonumber\\&+ \frac{ \Gamma_0^F}{\beta_0} \ln (1-2\chi_{0}) \ln \frac{  \mu_re^{\gamma_E}}{M_S},
\end{align}
with
\begin{align}\label{eq:chi}
\chi_{0} = \frac{\alpha_s(\mu_r) \beta_0}{4\pi}\ln\nu,
\end{align}
where $\beta_0=(11/3)C_A-(4/3)T_Fn_f$, $n_f$ is the number of active quark flavors and we choose $n_f = 3$. For the SU(3) color factors we have $T_F=1/2$, $C_F=4/3$ and $C_A=3$.
The presence of $\ln (1-2\chi_{0})$ in eq. \eqref{eq:sgfh} leads to the Landau singularity at the branch point
\begin{align}
\nu_1^L = \exp\biggr(\frac{2\pi}{\beta_0 \alpha_s(\mu_r)} \biggr).
\end{align}
The Landau singularity can be avoided by performing
following replacement in eqs. \eqref{eq:sgfh} and \eqref{eq:chi} as discussed in refs. \cite{Cacciari:2005uk,Shimizu:2005fp}
\begin{align}
\chi_0 \to \chi_{0\ast} = \frac{\alpha_s(\mu_r) \beta_0}{4\pi}\ln  \left(\frac{\nu}{1+\nu/\nu_{1\ast}^L}\right), \quad \nu_{1\ast}^L=\nu_1^L/a,
\end{align}
where $a$ is a parameter of order 1 but not smaller than 1.
Such a replacement prevents $\chi_0$ from entering the nonperturbative regime. And the nonperturbative effects would be compensated by the model $\tilde F^{\mathrm{mod}}[\COaSz](\nu)$. Finally, with the help of the numerical inverse Laplace transformation, we can transform eq. \eqref{eq:Resummed-SGD} back to momentum
space,
\begin{align}\label{eq:Resummed-SGD2}
F_{[\COaSz] \to \psi} (x,M_\psi,m_c,\mu_r) =&    \frac{1}{2\pi i}\int_{c-i \infty}^{c + i\infty} \mathrm{d} \nu e^{ (1/x-1) \nu} \tilde {F}_{[\COaSz] \to \psi}(\nu,M_\psi,m_c,  \mu_r).
\end{align}
The evolution of $\state{{3}}{P}{J,\lambda}{8}$ SGD can be solved similarly.

\subsection{The perturbative differential cross sections}\label{sec:cross-section}

We will calculate the perturbative differential cross sections of $e^+e^-$ annihilate to the $c\bar c$ pair in state $\COaSz$ and $\state{{3}}{P}{~}{8}$.
For P-wave channel, we only calculate the
polarization-summed hard part defined by
\begin{align}
H_{[\state{{3}}{P}{~}{8}]}^{(0)}(\hat z,M_\psi/x,s,x_{\mathrm{min}}/x,\mu_f)=\frac{1}{(d-1)^2}\sum_{J,\lambda} H_{[\state{{3}}{P}{J,\lambda}{8}]}^{(0)}(\hat z,M_\psi/x,s,x_{\mathrm{min}}/x,\mu_f).
\end{align}
We choose the $e^+e^-$ center-of-mass frame to perform the calculation.

\begin{figure}[ht]
\centering
\includegraphics[width=0.8\textwidth]{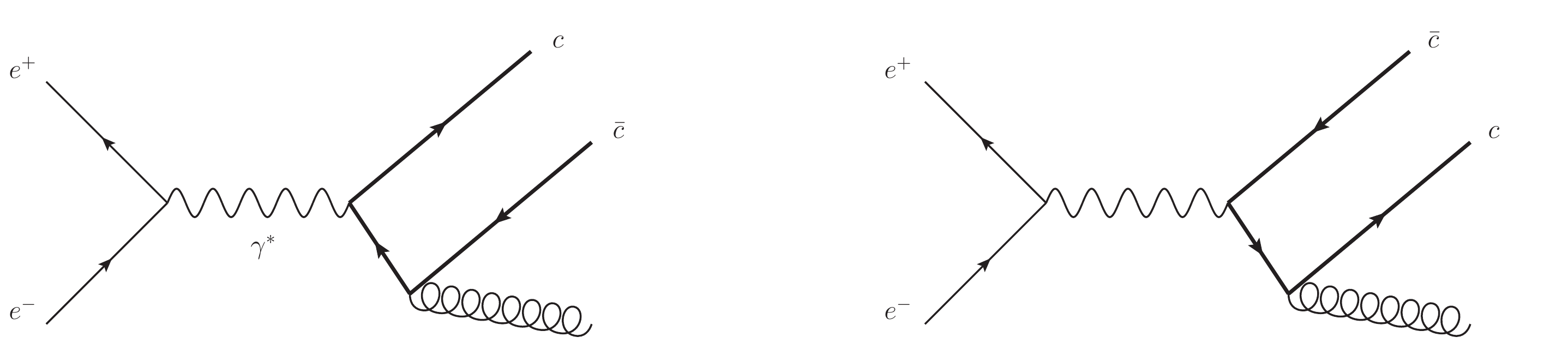}
\caption{\label{fig:LO} LO Feynman diagrams for $e^+e^- \to c\bar c(\COaSz,c \bar c[\state{{3}}{P}{~}{8}])+g$.}
\end{figure}

At LO, the Feynman diagrams are given in figure \ref{fig:LO}. Replacing the spinors of $c \bar c$ in the amplitude with the projectors in eq.~\eqref{eq:spinor-operator} and integrating over the solid angle of $q$ in the $c\bar c$ rest frame %by using the equations given in
, we can derive
\begin{equation}\label{eq:LO-S}
\frac{\mathrm{d}\sigma^{LO}_{S}}{\mathrm{d}z} ({z},M_\psi,s,m_c)
=
\frac{256 \pi ^2 \alpha ^2 \alpha _s  e_c^2 m_c^2 \mathcal{T} ^2  }{3
	s^2 M_{\psi}^3} \frac{1-r}{1+r}\delta(1- \bar{z}),
\end{equation}
for the $\COaSz$ channel. Here
\begin{equation}
\bar z = \frac{z}{1+r}, \qquad \mathcal{T}=\,_2F_1\left( \frac{1}{2} , 1 , \frac{3}{2} - \epsilon , \Delta^2  \right).
\end{equation}
For convenience,  we use $S$ and $P$ to represent the $c\bar c[\COaSz]$ and $c \bar c[\state{{3}}{P}{~}{8}]$ states, respectively. For the P-wave contribution, we list the result in appendix. \ref{ap:expression}.

\begin{figure}[ht]
\centering
\includegraphics[width=0.8\textwidth]{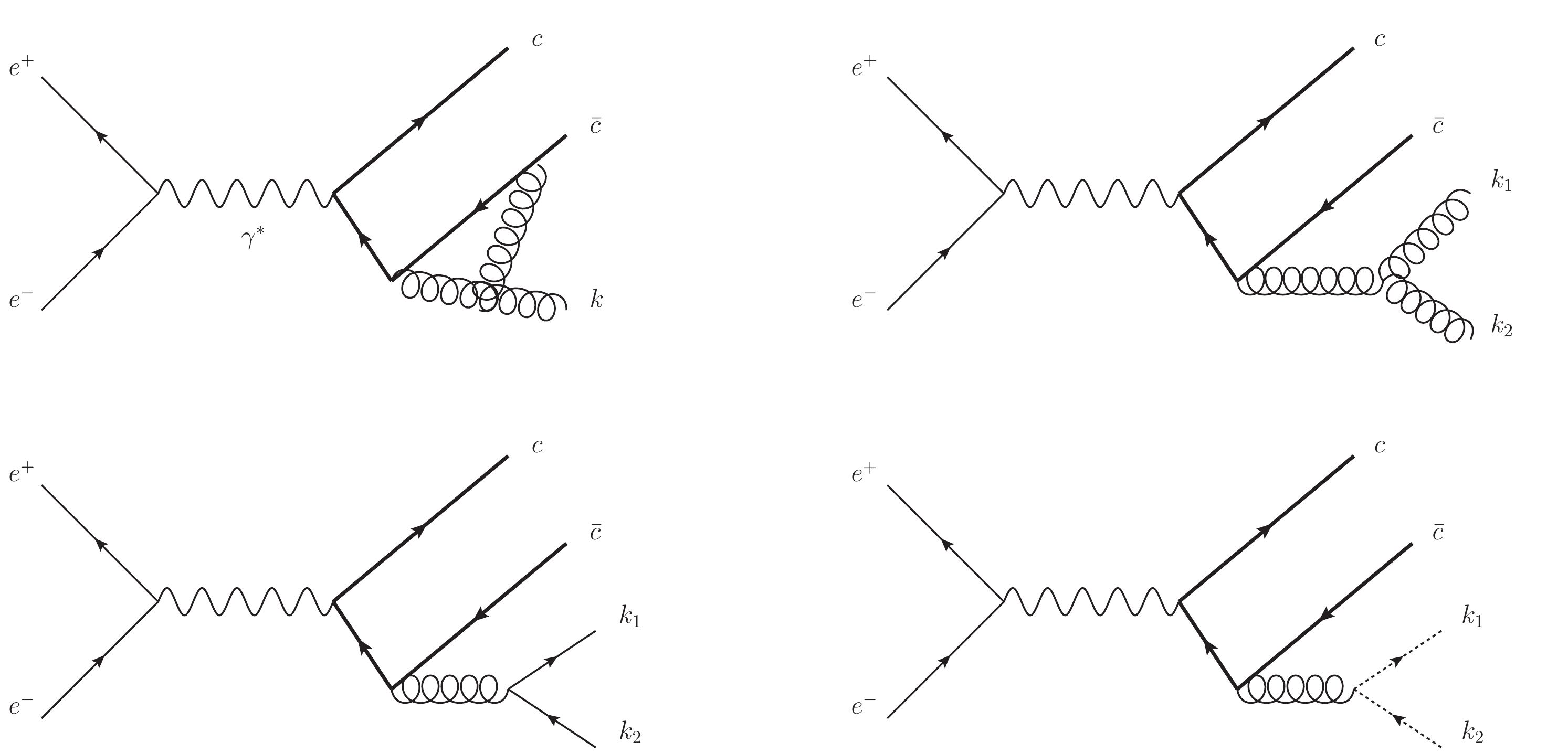}
\caption{\label{fig:NLO} Representative NLO Feynman diagrams for the $c\bar c[\COaSz]$ production from $e^+e^-$ annihilation.}
\end{figure}

We take the S-wave contribution as an example
to describe the calculation at NLO.  Some typical NLO Feynman diagrams are showed in figure \ref{fig:NLO}.
 The differential cross section is given by
\begin{equation}\label{eq:NLO-cross-section}
  \frac{\mathrm{d}\sigma^{NLO}_{S}}{\mathrm{d}z}
  =
  \frac{\mathrm{d}\sigma_{\mathrm{virtual}}}{\mathrm{d}z}+
  \frac{\mathrm{d}\sigma_{\mathrm{real}}}{\mathrm{d}z},
\end{equation}
with
\begin{subequations}
  \begin{align}
    \frac{\mathrm{d}\sigma_{\mathrm{virtual}}}{\mathrm{d}z}&=
     \frac{(d-2)e^2}{4(d-1)s^2}\int \mathrm{d}\Phi_2 \sum 2\mathrm{Re}\left(\mathcal{M}_{\mathrm{Born}}^*\mathcal{M}_{\mathrm{virtual}}\right)\\
    \frac{\mathrm{d}\sigma_{\mathrm{real}}}{\mathrm{d} z}&=
     \frac{(d-2)e^2}{4(d-1)s^2} \int \mathrm{d}\Phi_3 \sum|\mathcal{M}_{\mathrm{real}}|^2,
  \end{align}
\end{subequations}
where we have use Lorentz covariance and gauge invariance to relate the cross section in $e^+e^-$ annihilation to the decay rate of a virtual photon.
Therefore, $\mathcal{M}_{\mathrm{Born}}$ denotes the tree-level amplitude for $\gamma^\ast \to c\bar c(\COaSz)+g$, $\mathcal{M}_{\mathrm{virtual}}$ denotes
the one-loop amplitude for $\gamma^\ast \to c\bar c(\COaSz)+g$ and $\mathcal{M}_{\mathrm{real}}$ is the amplitude for real correction of $\gamma^\ast \to c\bar c(\COaSz)+X$, where $X$ can be two gluons, a light quark-antiquark pair or a ghost-ghost pair. $\sum$ means summation
over the polarizations final states and initial state virtual photon. $\mathrm{d}\Phi_{2(3)}$ means two-body (three-body) phase space which are given by
\begin{align}
  \mathrm{d}\Phi_2
  &=
  \frac{\mathrm{d}^{d - 1} \mathbf{P}_\psi}{(2 \pi)^{d - 1} 2 E_{\psi}}\frac{\mathrm{d}^{d - 1} \mathbf{k}}{(2 \pi)^{d - 1} 2 k^0}
  (2 \pi)^{d} \delta^{d}(q_\gamma-P_\psi-k) \delta(z - \frac{2E_\psi}{\sqrt{s}}),
  \\
   \mathrm{d} \Phi_{3}
  & =
  \frac{1}{N_s}
  \frac{\mathrm{d}^{d - 1} \mathbf{P}_\psi}{(2 \pi)^{d - 1} 2 E_{\psi}} \,
  \frac{\mathrm{d}^{d - 1} \mathbf{k}_{1}}{(2 \pi)^{d - 1} 2 k_1^0} \,
  \frac{\mathrm{d}^{d - 1} \mathbf{k}_{2}}{(2 \pi)^{d - 1} 2 k_2^0} \,
  (2 \pi)^{d} \delta^{d}(q_\gamma - P_\psi - k_{1} - k_{2})\delta(z - \frac{2E_\psi}{\sqrt{s}}),
\end{align}
where $q_\gamma$ is the momentum of $\gamma^\ast$, $k$ and $k_i$ ($i=1,2$) represent the momentum of emitted gluon, light quark or ghost, as showed in figure \ref{fig:NLO}, and $N_s$ is the symmetry factor for final-state particles.

In the calculation, we use FeynArts \cite{Hahn:2000kx} to generate  Feynman diagrams and amplitudes, and then use in-house code to contract Lorentz indices and carry out traces of Dirac matrices. According to eqs.~\eqref{eq:matching-relation},~\eqref{eq:lowest-order1}, we expand $m_c^2$ in the amplitudes around {$M_\psi^2/4$} and neglect the terms of $\mathcal {O}(\vert \textbf{q} \vert )$ before performing loop integration and phase space integration. We use the reverse unitary technique \cite{Anastasiou:2002yz,Anastasiou:2002qz,Anastasiou:2003yy} to transform the delta functions to propagator denominators,
\begin{equation}
\delta(x)
=
\frac{1}{2\pi}
\lim_{\varepsilon \to 0}
\left(
\frac{1}{x+i \varepsilon}
-
\frac{1}{x-i \varepsilon}
\right),
\end{equation}
then phase space integration can be treated similar as loop integration.
We use the integration-by-parts (IBP) method \cite{Chetyrkin:1981qh,Laporta:2001dd} and employ the package FIRE5 \cite{Smirnov:2014hma} to perform the reduction of loop integrals, which express cross sections as linear combinations of master integrals (MIs). To calculate these MIs,
we can set up differential equations with respect to $r$ \cite{Kotikov:1990kg,Gehrmann:1999as},
\begin{equation}\label{eq:DE}
  \frac{\partial}{\partial r}\mathbf{I}(r,\epsilon)=\mathbf{A}(r,\epsilon)\mathbf{I}(r,\epsilon).
\end{equation}
The boundary conditions at $r=1$ can be calculated  by using the method of region \cite{Beneke:1997zp}.
We then use the algorithm in \cite{Lee:2014ioa} to transform eq. \eqref{eq:DE} to the $\epsilon$-form \cite{Henn:2013pwa}. By solving the system, we   obtain the MIs which are expressed by Goncharov polylogarithms \cite{Goncharov:2001iea}. We further simplify the expressions by using the package PolyLogTools\cite{Duhr:2019tlz}. Combining the IBP coefficients and the MIs we obtain expressions of the virtual and real corrections.

The UV divergence in virtual correction can be removed by the renormalization. We choose the renormalization constants $Z_2$, $Z_m$, $Z_3$ that correspond the charm quark field, the charm quark mass and the gloun field respectively in the on-mass-shell (OS) scheme, and choose $Z_g$ that corresponds the QCD coupling in the minimal-subtractions ($\overline{\mathrm{MS}}$) scheme,
\begin{align}\label{eq: renDef}
  \delta Z_2^{\rm OS}&=-C_F\frac{\alpha_s}{4\pi}
  \left[\frac{1}{\epsilon_{\rm UV}}+\frac{2}{\epsilon_{\rm IR}}
  -3\gamma_E+3\ln\frac{4\pi\mu^2}{m_c^2}+4\right]+\mathcal
  {O}(\alpha_s^2),
  \nonumber\\
   \delta Z_m^{\rm OS}&=-3C_F\frac{\alpha_s}{4\pi}
  \left[\frac{1}{\epsilon_{\rm UV}}-\gamma_E+\ln\frac{4\pi\mu^2}{m_c^2}
  +\frac{4}{3}\right]+\mathcal
  {O}(\alpha_s^2), \nonumber\\
   \delta Z_3^{\rm OS}&=\frac{\alpha_s}{4\pi} (\beta_0-2C_A)
  \left[\frac{1}{\epsilon_{\rm UV}} -\frac{1}{\epsilon_{\rm IR}}\right]+\mathcal
  {O}(\alpha_s^2), \nonumber\\
    \delta Z_g^{\overline{\rm MS}}&=-\frac{\beta_0}{2}\,
    \frac{\alpha_s}{4\pi}
    \left[\frac{1}{\epsilon_{\rm UV}} -\gamma_E + \ln(4\pi)
    \right]+\mathcal
  {O}(\alpha_s^2).
\end{align}

Combining the virtual and real corrections, we get the differential cross section. The P-wave contribution can be calculated similarly.
Expressions of S-wave and P-wave contributions are given by
\begin{subequations}\label{eq:NLO-distribution}
\begin{align}
    \frac{\mathrm{d}\sigma_{S}^{NLO} }{\mathrm{d}z}
    =&  \sigma_{S}^{0}
      \Bigg\{
      \Big(\frac{\alpha_s}{4\pi} R_{S}(M_\psi,r,\mu_r)- \frac{\alpha_s}{4\pi} \frac{1}{C_A}\frac{\pi^2}{\Delta} \Big) \delta(1-\bar{z})
      \nonumber\\&\hspace{-1cm}
      + \frac{\alpha_s}{4\pi}
              \Bigg[
                  \Bigg(
                     4C_A\ln \frac{\left(\sqrt{(r+1)^2 \bar{z}^2-4 r}+r \bar{z}-2 r+\bar{z}\right)^2}{4 r (r+1)}-P[r,\bar{z}](4C_A+\beta_0)
                  \Bigg)
                  \nonumber\\&\hspace{-1cm}
                  \times \left[\frac{1}{1-\bar{z}}\right]_+
                  -
                  4C_A \left[\frac{\ln(1-\bar{z})}{1-\bar{z}}\right]_+ +P[r,\bar{z}]\mathcal {R}_{S}(\bar z,M_\psi,r,\mu_r)
              \Bigg]
      \Bigg\}  + \mathcal {O}(\Delta),\\
\frac{\mathrm{d}\sigma_{P}^{NLO} }{\mathrm{d}z}
    =&  \sigma_{P}^{0}
      \Bigg\{
      \Big(\frac{\alpha_s}{4\pi} R_{P}(M_\psi,r,\mu_r)- \frac{\alpha_s}{4\pi} \frac{1}{C_A}\frac{\pi^2}{\Delta} \Big)\delta(1-\bar{z})
      \nonumber\\&\hspace{-1cm}
      + \frac{\alpha_s}{4\pi}
              \Bigg[
                  \Bigg(  4C_A \ln
                  \frac{ \left(\sqrt{(r+1)^2 \bar{z}^{2}-4 r}+r \bar{z}-2 r+\bar{z} \right)^2}{4 r (r+1)}
                   -P[r,\bar{z}](4C_A+\beta_0)
                  \Bigg)\nonumber\\&\hspace{-1cm}
                     \times \left[\frac{1}{1-\bar{z}}\right]_+
                  -
                  4C_A \left[\frac{\ln(1-\bar{z})}{1-\bar{z}}\right]_+ +P[r,\bar{z}]\mathcal {R}_{P}(\bar z,M_\psi,r,\mu_r)
              \Bigg]
      \Bigg\}  + \mathcal {O}(\Delta),
\end{align}
\end{subequations}
where
\begin{equation}
\sigma_{S}^{0}
=
\frac{64 \pi ^2 \alpha^2 \alpha _s  e_c^2  }{3
	s^2 M_{\psi}}\frac{1-r}{1+r}, \quad\quad  \sigma_{P}^{0}
=
\frac{256 \pi ^2 \alpha^2 \alpha _s  e_c^2   }{27
	s^2 M_{\psi}^3}\frac{3+2r+7r^2}{(1-r)(1+r)},
\end{equation}
$\mu_r$ is the renormalization scale and the function $P[r,{\bar{z}}]$ is defined as
\begin{align}\label{eq:pfac}
  P[r,{\bar{z}}] = \frac{\sqrt{(1+r)^2 {\bar{z}}^2-4 r}}{1-r}.
\end{align}
The expressions of $R_S$, $R_P$, $\mathcal {R}_S$ and $\mathcal {R}_P$ are listed in appendix \ref{ap:expression}.
Our analytical result for the S-wave differential cross section is equivalent to that in \cite{Sun:2018yam}, and the analytical result for the P-wave channel is new.

\subsection{Matching the short-distance hard parts}

By inserting eqs. \eqref{eq:LO-S}, \eqref{eq:LO-P}, \eqref{eq:NLO-distribution}, \eqref{eq:SGDresult} into the matching relation eq. \eqref{eq:matching-relation} and using eq. \eqref{eq:lowest-order1}, we can derive the short-distance hard parts $H^{(0)}$ directly. At LO we have
\begin{subequations}\label{eq:LO-hard-part}
  \begin{align}
  H_{[\COaSz]}^{(0),LO}(z,M_\psi,s,x_{\mathrm{min}},\mu_f)=& \frac{64 \pi ^2 \alpha^2 \alpha _s  e_c^2  }{3
	s^2 M_{\psi}}\frac{1-r}{1+r}\delta(1 - \bar z),
\\
 H_{[\state{{3}}{P}{~}{8}]}^{(0),LO}(z,M_\psi,s,x_{\mathrm{min}},\mu_f)=& \frac{256 \pi ^2 \alpha^2 \alpha _s  e_c^2   }{27
	s^2 M_{\psi}^3}\frac{3+2r+7r^2}{(1-r)(1+r)} \delta(1 - \bar z).
  \end{align}
\end{subequations}
One can find the short-distance hard part in P-wave channel is plagued with a singularity associated with the limit $r=M_\psi^2/s\to 1$. Inserting the above hard parts into the SGF formula, $M_\psi$ will be replaced by $M_\psi/x$, and thus such singularity appears at the threshold limit $M_{c\bar c}^2=M_\psi^2/x^2 \to s$, or $x\to \sqrt{r}$. In the threshold limit, the emitted gluon at LO in  figure~\ref{fig:LO} is very soft and its effect should be included in nonperturbative SGDs rather than in short-distance hard part.
In fact, in the threshold region, $\state{{3}}{S}{1}{1}$  channel can also contribute via $e^+ e^- \to \gamma^*\to c\bar c(\state{{3}}{S}{1}{1})$, followed by the hadronization process described by the SGD $F_{[\state{{3}}{S}{1}{1}] \to \psi }$. When  $\state{{3}}{S}{1}{1}$  channel is included, the matching process to determine  the $\state{{3}}{P}{~}{8}$ coefficient will have a contribution from the perturbative transition $F_{[\state{{3}}{S}{1}{1}] \to c\bar c[\state{{3}}{P}{~}{8}] }$, which will cancel the aforementioned threshold singularity.

However, in this paper we will be only interested in B-factories where $r\approx0.1$ is very small. Then, when $x\sim \sqrt{r}$, the nonperturbative quantities SGDs have significant perturbatively calculable effects, which can be avoided by introducing a reasonable $x_{\textrm{min}}$. By choosing
$x_{\mathrm{min}}>\sqrt{r}$, both the contribution from  $e^+ e^- \to \gamma^*\to c\bar c(\state{{3}}{S}{1}{1})$ and the threshold singularity of P-wave channel will disappear.  This is consistent with the fact that SGF formula is insensitive to small $x_{\textrm{min}}$.

%
%in this threshold limit the $c\bar c$ pair can also be in $\state{{3}}{S}{1}{1}$ state. At tree level, it correspond to the process $\gamma^*\to c\bar c(\state{{3}}{S}{1}{1})$. Thus when matching the $\state{{3}}{P}{~}{8}$ coefficient with $x_{\mathrm{min}}=0$, we also need to include the effect of $F_{[\state{{3}}{S}{1}{1}] \to c\bar c[\state{{3}}{P}{~}{8}] }$, which will cancel the divergence. This is also evidence that the $x_{\textrm{min}}$ dependence of the theory should be weak. In our case the threshold limit $x\to \sqrt{r}$ is too small, at which intermediate $c\bar c$ pair will emit hard gluons during the hadronization process, which is perturbative. Therefore, introducing a cutoff $x_{\textrm{min}}$ to prevent $x$ from entering the perturbative regime is reasonable in physics.
%
%This singularity can be avoided if the introduced cutoff satisfies $x_{\mathrm{min}}>\sqrt{r}$. It should be noted that the introduction of $x_{\textrm{min}}$ is not a must. If we choose $x_{\mathrm{min}}=0$, the singularity
%can be subtracted by the SGD $F_{[\state{{3}}{S}{1}{1}] \to \psi }$. The divergence appears at the threshold limit $M_{c\bar c}^2=M_\psi^2/x^2 \to s$, with $M_{c\bar c}$ denotes the mass of intermediate $c\bar c$ pair, and it can be regularized by dimensional regularization.

At NLO, we have
  \begin{align}
  H_{[\COaSz]}^{(0),NLO}(z,M_\psi,s,x_{\mathrm{min}},\mu_f)
    =& \biggr[\frac{\mathrm{d}\sigma_{S}^{NLO} }{\mathrm{d}z}- \int_{\mathrm{max}[\frac{\sqrt{z^2-4r}+z}{2},
   x_{\mathrm{min}}]}^1
    \frac{\mathrm{d}x}{x}
\frac{256 \pi ^2 \alpha ^2 \alpha _s  e_c^2  m_c^2\mathcal{T}_x^2 x^3 }{3
	s^2 M_{\psi}^3}
\nonumber\\
      &\hspace{-3.5cm}\times
       \left(1-\frac{r}{x^2}\right) \delta(1+r/x^2 - z/x)
    F^{NLO}_{{[\COaSz]}\to {c\bar c[\COaSz]}}(x,M_\psi,m_c,\mu_f)\biggr]\biggr\vert_{m_c=M_\psi/2},
  \end{align}
where
$\mathcal{T}_x=\mathcal{T}\vert_{M_\psi\to M_\psi/x}$. By integrating over $x$, we then obtain
  \begin{align}\label{eq:NLO-hardpart-S}
  H_{[\COaSz]}^{(0),NLO}(z,M_\psi,s,x_{\mathrm{min}},\mu_f)
    =&\frac{\sigma_{S}^{0}\alpha_s}{4\pi} \
      \Bigg\{
      \biggr( R_{S}(M_\psi,r,\mu_r) +  C_A \Big( \ln^2\frac{\mu_f^2 e^{ - 1}}{M_\psi^2} +\frac{\pi^2}{6} -1 \Big) \biggr)
      \nonumber\\
      &\hspace{-4cm}\times \delta(1-{\bar{z}})
      + \biggl(
      {4C_A}\ln \frac{\left(\sqrt{(r+1)^2 {\bar{z}}^2-4 r}+r {\bar{z}}-2 r+{\bar{z}}\right)^2}{4 r (r+1)}-P[r,{\bar{z}}](4C_A+\beta_0)
                  \biggr)
     \nonumber\\
                     &\hspace{-4cm}\times \left[\frac{1}{1-{\bar{z}}}\right]_+
                  -
                  {4C_A} \left[\frac{\ln(1-{\bar{z}})}{1-{\bar{z}}}\right]_+ +P[r,{\bar{z}}]\mathcal {R}_{S}(\bar z,M_\psi,r,\mu_r)
       - \theta \left(z \geq \frac{x_{\mathrm{min}}^2+r}{x_{\mathrm{min}}} \right)
       \nonumber\\
                     &\hspace{-4cm}\times \frac{z(1+r) x_1^4
                     \mathcal {T}^{\prime 2}C_A}{(1-r)(x_1^2+r)}
  \biggr[ \frac{4x_1}{(1-x_1)_+}  \ln\biggr(\frac{ x_1^2\mu_f^2 e^{ - 1}}{M_\psi^2}\biggr)
 - 8x_1
 \biggr( \frac{\ln(1-x_1)}{1-x_1} \biggr)_+
 \biggr] \Bigg\},
  \end{align}
 with $\theta(t)=0$ ($1$) if $t$ is false (true),  and
 \begin{subequations}
 \begin{align}
 \mathcal {T}^{\prime }=&\frac{1}{2\sqrt{1-x_1^2}}\ln\frac{1+\sqrt{1-x_1^2}}{1-\sqrt{1-x_1^2}},\\
x_1 =& \frac{\sqrt{z^2-4 r }+z}{2 }.
\end{align}
\end{subequations}
We can find the above result still contains large logarithms at the limit $z\to 1+r$. These remained large logarithms originate from the collinear radiations recoil against the $c \bar c$ pair in the threshold region, which are not factorized in pure SGF. They can be resummed by introduce a jet function, as we will explain later.

Similarly, for the P-wave contribution we have
\begin{align}\label{eq:NLO-hardpart-P}
  H_{[\state{{3}}{P}{~}{8}]}^{(0),NLO}(z,M_\psi,s,x_{\mathrm{min}},\mu_f)
    =&\frac{\sigma_{P}^{0}\alpha_s}{4\pi} \
      \Bigg\{
      \biggr( R_{P}(M_\psi,r,\mu_r) +  C_A \Big( \ln^2\frac{\mu_f^2 e^{ - 1}}{M_\psi^2} +\frac{\pi^2}{6} -1 \Big) \biggr)
      \nonumber\\
      &\hspace{-4cm}\times \delta(1-{\bar{z}})
      + \biggl(
      {4C_A}\ln \frac{\left(\sqrt{(r+1)^2 {\bar{z}}^2-4 r}+r {\bar{z}}-2 r+{\bar{z}}\right)^2}{4 r (r+1)}-P[r,{\bar{z}}](4C_A+\beta_0)
                  \biggr)
     \nonumber\\
                     &\hspace{-4cm}\times \left[\frac{1}{1-{\bar{z}}}\right]_+
                  -
                  {4C_A} \left[\frac{\ln(1-{\bar{z}})}{1-{\bar{z}}}\right]_+ +P[r,{\bar{z}}]\mathcal {R}_{P}(\bar z,M_\psi,r,\mu_r)
       - \theta \left(z \geq \frac{x_{\mathrm{min}}^2+r}{x_{\mathrm{min}}} \right)
       \nonumber\\
                     &\hspace{-4cm}\times \frac{z\Theta(x_1,r)
                     C_A}{z^2-4r}
  \biggr[ \frac{4x_1}{(1-x_1)_+}  \ln\biggr(\frac{ x_1^2\mu_f^2 e^{ - 1}}{M_\psi^2}\biggr)
 - 8x_1
 \biggr( \frac{\ln(1-x_1)}{1-x_1} \biggr)_+
 \biggr] \Bigg\},
\end{align}
with
\begin{align}
\Theta(x_1,r)=&\frac{ (1-r^2 ) x_1^2}{4 (7 r^2 + 2 r + 3) (x_1 - 1)^2 (x_1 + 1)^4 ( x_1^2+r )} \biggr[ r^2 \Big(27 \mathcal {T}^{\prime 2} x_1^8 +
      72 \mathcal {T}^{\prime 2} x_1^7
\nonumber\\
&+ (72 \mathcal {T}^{\prime 2} - 264 \mathcal {T}^{\prime} + 79) x_1^4 +
      24 (3 \mathcal {T}^{\prime 2} - 11 \mathcal {T}^{\prime} + 7) x_1^3
\nonumber\\
& +
2 (18 \mathcal {T}^{\prime 2} - 60 \mathcal {T}^{\prime} + 65) x_1^2 + 6 \mathcal {T}^{\prime} (15 \mathcal {T}^{\prime} - 11) x_1^6 +
      24 \mathcal {T}^{\prime} (3 \mathcal {T}^{\prime} - 7) x_1^5
\nonumber\\
& + 48 x_1 + 16 \Big) +
   4 r (x_1 - 1) x_1^2 \Big(27 \mathcal {T}^{\prime 2} x_1^4 + 4 (9 \mathcal {T}^{\prime 2} - 9 \mathcal {T}^{\prime} + 2) x_1^3
\nonumber\\
&   +
      2 (9 \mathcal {T}^{\prime 2} - 9 \mathcal {T}^{\prime} - 11) x_1^2 +
      6 \mathcal {T}^{\prime} (3 \mathcal {T}^{\prime} - 1) x_1^5 + (6 \mathcal {T}^{\prime} - 26) x_1 - 5 \Big)
\nonumber\\
&   +
   6 x_1^4 \Big(3 \mathcal {T}^{\prime 2} x_1^6 + 6 \mathcal {T}^{\prime 2} x_1^5 +
      2 (3 \mathcal {T}^{\prime 2} - 5 \mathcal {T}^{\prime } + 1) x_1^4 + (6 \mathcal {T}^{\prime 2} - 20 \mathcal {T}^{\prime} + 9) x_1^2
\nonumber\\
&   + (4 -
         16 \mathcal {T}^{\prime}) x_1^3 + (4 \mathcal {T}^{\prime} + 2) x_1 + 4 \Big)\biggr].
\end{align}
{We can find in the last line of eq. \eqref{eq:NLO-hardpart-P} that the singularity at $z\to 2\sqrt{r}$ is avoided due to the introduction of cut off $x_{\mathrm{min}}>\sqrt{r}$. As a result, the cross section $\sigma_{J/\psi}$ is $x_{\mathrm{min}}$ dependent. However, as demonstrated in appendix~\ref{ap:cutoff-dependence}, for small and moderate $x_{\textrm{min}}$, the $x_{\textrm{min}}$ dependence of $\sigma_{J/\psi}$ is either $\alpha_s$ suppressed or $\Lambda_{\mathrm{QCD}}/M_\psi$ suppressed. Thus the cross section $\sigma_{J/\psi}$ in SGF is insensitive to the choice of $x_{\mathrm{min}}$.}

Inserting eqs. \eqref{eq:LO-hard-part}, \eqref{eq:NLO-hardpart-S} and \eqref{eq:NLO-hardpart-P} into eq. \eqref{eq:v-expansion}, we then obtain the differential cross section at the lowest order in the velocity expansion
 \begin{align}\label{eq:lowest-order}
 \frac{\mathrm{d}\sigma_{J/\psi}}{\mathrm{d}z}
 =& \int_{\mathrm{max}[(z+\sqrt{z^2-4r})/2,x_{\mathrm{min}}]}^1 \frac{\mathrm{d}x}{x}   \biggr[H_{[\COaSz]}^{(0)}(\hat z,M_\psi/x,s,x_{\mathrm{min}}/x,\mu_f)
 F_{[\COaSz] \to \psi }(x,M_{\psi},m_c, \mu_f) \nonumber\\
 &+\sum_{J,\lambda} H_{[\state{{3}}{P}{~}{8}]}^{(0)}(\hat z,M_\psi/x,s,x_{\mathrm{min}}/x,\mu_f)
 F_{[\state{{3}}{P}{J,\lambda}{8}] \to \psi }(x,M_{\psi},m_c, \mu_f)\biggr].
\end{align}

\section{NRQCD factorization}\label{sec:NRQCD}

As a comparison, we also present here the results of NRQCD factorization. The CO contribution for inclusive $J/\psi$ production can be factorized as
  \begin{equation}\label{eq:nrqcd-CO}
  \mathrm{d}\sigma(e^+e^- \to J/\psi + X)
  =
  \mathrm{d}\hat{\sigma}[\COaSz]
  \langle  \mathcal{O}^{J/\psi}(\COaSz)  \rangle +   \frac{1}{9}\mathrm{d}\hat{\sigma}[\state{{3}}{P}{~}{8}]
  \langle  \mathcal{O}^{J/\psi}(\state{{3}}{P}{~}{8})  \rangle,
\end{equation}
where the factor $1/9$ is to average over polarizations of intermediate states, and $\langle  \mathcal{O}^{J/\psi}(\state{{3}}{P}{~}{8})  \rangle$ is the polarization-summed LDME defined by
\begin{equation}
  \langle  \mathcal{O}^{J/\psi}(\state{{3}}{P}{~}{8})  \rangle
  =
  \sum_{J,\lambda}\langle  \mathcal{O}^{J/\psi}(\state{{3}}{P}{J,\lambda}{8})  \rangle,
\end{equation}
which has following relation
at the lowest order in velocity approximation,
\begin{equation}\label{eq:P-LDME}
  \langle  \mathcal{O}^{J/\psi}(\state{{3}}{P}{~}{8})  \rangle
  \approx
  9\langle \mathcal{O}^{J/\psi}[^3P_0^{[8]}] \rangle.
\end{equation}

Based on perturbative differential cross sections calculated in section. \ref{sec:cross-section}, we can easily obtain the SDCs for NRQCD factorization
\begin{subequations}\label{eq:SDCs}
\begin{align}
    \frac{\mathrm{d}\hat{\sigma}[\COaSz] }{\mathrm{d}z}
    =& \ \sigma_{S}^{\prime 0} \
      \Bigg\{
      \Big(1+\frac{\alpha_s}{4\pi} R_{S}(2m_c,r^\prime,\mu_r)\Big)\delta(1-z^\prime)
      \nonumber\\
      &\hspace{-1cm}
      + \frac{\alpha_s}{4\pi}
              \Bigg[
                  \Biggl(
                     {4C_A}\ln \frac{\left(\sqrt{(r^\prime+1)^2 z^{\prime2}-4 r^\prime}+r^\prime z^\prime-2 r^\prime+z^\prime\right)^2}{4 r^\prime (r^\prime+1)}-P[r^\prime,z^\prime](4C_A+\beta_0)
                  \Biggr)
                  \nonumber\\
      &\hspace{-1cm}
      \times \left[\frac{1}{1-z^\prime}\right]_+
                  -
                  {4C_A} \left[\frac{\ln(1-z^\prime)}{1-z^\prime}\right]_+ +P[r^\prime,z^\prime]\mathcal {R}_{S}(z^\prime,2m_c,r^\prime,\mu_r)
              \Bigg]
      \Bigg\}+\mathcal
  {O}(\alpha_s^3),\\
\frac{\mathrm{d}\hat{\sigma}[\state{3}{P}{~}{8}] }{\mathrm{d}z}
    =& \ \sigma_{P}^{\prime 0} \
      \Bigg\{
      \Big(1+\frac{\alpha_s}{4\pi} R_{P}(2m_c,r^\prime,\mu_r)\Big)\delta(1-z^\prime)
      \nonumber\\
      &\hspace{-1cm}
      + \frac{\alpha_s}{4\pi}
              \Bigg[
                  \Biggl(
                     {4C_A}\ln \frac{\left(\sqrt{(r^\prime+1)^2 z^{\prime2}-4 r^\prime}+r^\prime z^\prime-2 r^\prime+z^\prime\right)^2}{4 r^\prime (r^\prime+1)}-P[r^\prime,z^\prime](4C_A+\beta_0)
                  \Biggr)
                  \nonumber\\
      &\hspace{-1cm}
      \times \left[\frac{1}{1-z^\prime}\right]_+
                  -
                  {4C_A} \left[\frac{\ln(1-z^\prime)}{1-z^\prime}\right]_+ +P[r^\prime,z^\prime]\mathcal {R}_{P}(z^\prime,2m_c,r^\prime,\mu_r)
              \Bigg]
      \Bigg\}+\mathcal
  {O}(\alpha_s^3),
  \end{align}
  \end{subequations}
where
\begin{equation}
r^\prime=\frac{4m_c^2}{s}, \quad \quad z^\prime = \frac{z}{1+r^\prime},
\end{equation}
and
\begin{equation}
\sigma_{S}^{\prime0}
=
\frac{32 \pi ^2 \alpha^2 \alpha _s  e_c^2  }{3
	s^2 m_c}\frac{1-r^\prime}{1+r^\prime}, \quad\quad  \sigma_{P}^{\prime0}
=
\frac{32 \pi ^2 \alpha^2 \alpha _s  e_c^2   }{3
	s^2 m_c^3}\frac{3+2r^\prime+7r^{\prime2}}{(1-r^\prime)(1+r^\prime)}.
\end{equation}
Using the same multi-loop calculation techniques, we can also obtain integrated cross sections. The corresponding SDCs are given by
\begin{subequations}\label{eq:integrated-SDCs}
\begin{align}
    \hat{\sigma}[\COaSz]
    =&(1+r^\prime)\sigma_{S}^{\prime 0} \Big(1+ \frac{\alpha_s}{4\pi} \mathcal {G}_{S}(2m_c,r^\prime,\mu_r)\Big)+\mathcal
  {O}(\alpha_s^3),\\
  \hat{\sigma}[\state{3}{P}{~}{8}]
    =&(1+r^\prime)\sigma_{P}^{\prime 0} \Big(1+ \frac{\alpha_s}{4\pi} \mathcal {G}_{P}(2m_c,r^\prime,\mu_r)\Big)+\mathcal
  {O}(\alpha_s^3),
\end{align}
\end{subequations}
where $\mathcal {G}_S$ and $\mathcal {G}_P$ are given in appendix \ref{ap:expression}. Our results can reproduce that in ref. \cite{Zhang:2009ym} if we choose the same parameters therein.

Clearly, at the endpoint region, $z\to 1+r^\prime$ (i.e. $ z^\prime\to 1$), the fixed-order results of the SDCs in eq. \eqref{eq:SDCs} suffer from large threshold logarithms. In order not to spoil the convergence of perturbative expansion, these threshold logarithms have to be resummed to all orders.
Such resummation of the threshold logarithms to LO+NLL accuracy has been studied in ref. \cite{Fleming:2003gt} within the SCET framework. According to \cite{Fleming:2003gt,Sun:2018yam}, at the endpoint region we have following factorization formula
\begin{align}\label{eq:scet}
  \frac{\mathrm{d}\hat\sigma[n]}{\mathrm{d}  z^\prime}\biggr\vert_{\mathrm{endpoint}}
  =
  P[r^\prime,z^\prime](1+r^\prime)
  \sigma_{[n]}^{\prime 0}
  \mathcal {H}^{[n]}(\mu_r,\mu)
  \int_{z^\prime}^1\mathrm{d}\xi \mathcal {S}^{[n]}(1-\xi,\mu)\mathcal {J}(s(1+r)(\xi- z^\prime),\mu),
\end{align}
where a phase space factor $P[r^\prime,z^\prime]$ is introduced \cite{Fleming:2003gt} with $P[r^\prime,1]=1$.
Fixed-order expression of the SDCs at the endpoint can be read from eq. \eqref{eq:SDCs},
{
\begin{align}\label{eq:endpoint}
\frac{\mathrm{d}\hat\sigma[n]}{\mathrm{d}  z^\prime}\biggr\vert_{\mathrm{endpoint}}
 =& P[r^\prime,z^\prime](1+r^\prime)
  \sigma_{[n]}^{\prime 0} \Biggr\{ \Big(1+\frac{\alpha_s}{4\pi} R_{[n]}(2m_c,r^\prime,\mu_r)\Big) \delta(1-z^\prime)
\nonumber\\
      &\hspace{-2cm}
 + \frac{\alpha_s}{4\pi} \biggr[  \biggr( 4C_A \ln\frac{(1-r^\prime)^2}{r^\prime(1+r^\prime)}-4C_A - \beta_0 \biggr)\left(\frac{1}{1-z^\prime} \right)_+
  - 4C_A \left(\frac{\ln(1-z^\prime)}{1-z^\prime} \right)_+ \biggr] \Biggr\} + \mathcal
  {O}(\alpha_s^3).
\end{align}}
The shape function $\mathcal {S}^{[n]}(\xi)$ are defined in terms of ultrasoft fields that carry $\mathcal{O}(\Lambda_{\mathrm{QCD}})$ momentum
\begin{subequations}
  \begin{align}
    \mathcal {S}^{[\COaSz]}(1-\xi,\mu)=&
   \frac{
      \langle 0|
      \chi^\dagger T^a\psi
      \,a_{\psi}^{\dagger}a_{\psi}\,
      \delta((1-\xi) + i \hat l\cdot D)
      \psi^\dagger T^a\chi
      |0 \rangle
    }{4m_c\langle \mathcal{O}^{J/\psi}[\COaSz] \rangle},\\
    \mathcal {S}^{[\COcPz]}(1-\xi,\mu)=&
    \frac{\frac{1}{3} \langle 0|
      \chi^\dagger \left( -\frac{i}{2}\overleftrightarrow{\mathbf{D}} \cdot \sigma \right) T^a\psi
      \,a_{\psi}^{\dagger}a_{\psi}\,
      \delta((1-\xi) + i \hat l\cdot D)
      \psi^\dagger \left( -\frac{i}{2}\overleftrightarrow{\mathbf{D}} \cdot \sigma \right) T^a\chi
      |0 \rangle
    }{4m_c\langle \mathcal{O}^{J/\psi}[\COcPz] \rangle},
  \end{align}
\end{subequations}
where the ultrasoft covariant derivative is given as $D^\mu=\partial^\mu-ig_s A^{\mu}_{us}$, the lightlike vector $\hat l^\mu$ is defined as $\hat l^\mu=\sqrt{2}l^\mu/M_S$ with $M_{S} \equiv 2m_c (1+r)/(1-r)$, and $\psi$ and $\chi$ denote the Pauli spinor fields in NRQCD that annihilates a heavy quark and creates a heavy antiquark, respectively.
The jet function describes the collinear radiations recoil against the $J/\psi$ in the threshold region, which is defined as
\begin{equation}
  \mathcal {J}(p^2,\mu) = -\frac{s(1+r)}{4\pi}{\rm Im} \left[i\int d^4y e^{ip\cdot y}
  \langle0|T\left(\mathrm{Tr}[T^a B_\perp^{(0)\beta}(y)]
  \mathrm{Tr}[T^a B_{\perp \beta}^{(0)}(0)]\right)|0\rangle\right],
\end{equation}
where the superscript $(0)$ denotes the bare field, $B_{\perp}^\mu$ is the
collinear gauge invariant effective field \cite{Fleming:2003gt}, and the factor $s(1+r)$ is chosen to provide a convenient normalization for the process considered here.
According to \cite{Bauer:2001rh,Becher:2009th}, the fixed-order expression of the shape function and jet function are given by
\begin{subequations}\label{eq:shape-jet-function}
\begin{align}
    \mathcal {S}^{[n]}(1-\xi,\mu) =& \delta(1-\xi) + \frac{\alpha_s C_A} {4\pi}
      \biggr[ \frac{4}{(1-\xi)_+}  \ln\biggr(\frac{ \mu^2 e^{ - 1}}{M_{S}^2}\biggr)
     - 8
     \biggr( \frac{\ln(1-\xi)}{1-\xi} \biggr)_+
    \nonumber\\
    &- \delta(1-x)  \biggr( \ln^2\biggr(\frac{\mu^2 e^{ - 1}}{M_{S}^2}\biggr) +\frac{\pi^2}{6} -1 \biggr)  \biggr] + \mathcal
  {O}(\alpha_s^2) ,\\
  \mathcal {J}(M_{J}^2(\xi -z^\prime ),\mu) =& \delta(\xi -z^\prime ) + \frac{\alpha_s} {4\pi}
  \biggr\{
  \biggr[C_A\biggr(\frac{67}{9} - \pi^2\biggr)- \frac{20}{9} T_fn_f  - \beta_0 \ln\frac{M_{J}^2}{\mu^2}
  \nonumber\\
  &+ 2C_A   \ln^{2}\frac{M_{J}^2}{\mu^2}   \biggr] \delta(\xi -z^\prime )
  +\biggr(4C_A \ln\frac{M_{J}^2}{\mu^2} - \beta_0\biggr) \biggr[ \frac{1}{\xi -z^\prime}\biggr]_+
  \nonumber\\
  &
  +  4C_A   \biggr[ \frac{\ln (\xi -z^\prime)}{\xi -z^\prime }\biggr]_+
  \biggr \} + \mathcal
  {O}(\alpha_s^2),
\end{align}
\end{subequations}
where $M_{J}^2 \equiv s(1+r)$.

Similar to SGF, to perform the resummation of the factorization formula eq. \eqref{eq:scet}, we transform it to Laplace space
\begin{align}\label{eq:Laplace}
  \tilde{\sigma}[n](\nu)=& \int_0^\infty \mathrm{d} \omega e^{-\nu\omega} \frac{1}{ P[r^\prime,z^\prime]}\frac{\mathrm{d}\hat\sigma[n]}{\mathrm{d} z^\prime}\biggr\vert_{\mathrm{endpoint}}
  \nonumber\\
  =&(1+r^\prime)\sigma_{[n]}^{\prime 0}
  \mathcal {H}^{[n]}(\mu_r,\mu) \int_0^\infty \mathrm{d} \omega  e^{-\nu\omega}
  \int_0^{\omega}\mathrm{d}\omega^\prime \mathcal {S}^{[n]}(\omega^\prime,\mu)\mathcal {J}(M_{J}^2(\omega-\omega^\prime),\mu)
  \nonumber\\
  =&
  (1+r^\prime)\sigma_{[n]}^{\prime 0}
  \mathcal {H}^{[n]}(\mu_r,\mu)
  \tilde {\mathcal {S}}^{[n]}(\nu,\mu)
  \tilde {\mathcal {J}}(\nu,\mu),
\end{align}
where we introduce $\omega=1-z^\prime$ and $\omega^\prime = 1-\xi$, with $\omega^\prime$ denoting the longitudinal momentum fraction of the
emitted ultrasoft gluons in the shape function. Different from the momentum fraction $\omega_0=(1-x)/x$ in SGD, which takes values between $0$ and $\infty$, $\omega^\prime$ takes values between $0$ and $1$. Actually, the exact form of $\omega^\prime$ should be $\omega^\prime = (1-\xi)/\xi$, which also runs from $0$ to $\infty$. However, in NRQCD+SCET approach one only consider the resummation in the endpoint region ($\xi\to 1$), where one has $\omega^\prime = (1-\xi)/\xi \sim 1-\xi$, i.e. the evolution of the terms at higher order in $1-\xi$ is not included. While in the evolution of SGD, these contributions are included. In eq. \eqref{eq:Laplace}, $\tilde {\mathcal {S}}^{[n]}(\nu,\mu)$ and $\tilde {\mathcal {J}}(\nu,\mu)$ are the Laplace transformation of shape function and jet function,
\begin{subequations}
\begin{align}
  \tilde {\mathcal {S}}^{[n]}(\nu,\mu) =&\int_0^\infty \mathrm{d} \omega^\prime e^{-\nu\omega^\prime} \mathcal {S}^{[n]}(\omega^\prime,\mu),\\
  \tilde {\mathcal {J}}(\nu,\mu)=& \int_0^\infty \mathrm{d} \omega^\prime e^{-\nu\omega^\prime} \mathcal {J}(M_{J}^2\omega^\prime,\mu).
\end{align}
\end{subequations}
The components $\tilde {\mathcal {S}}^{[n]}$, $\tilde {\mathcal {J}}$ and $\mathcal {H}^{[n]}$ satisfy the renormalization group equations
\begin{subequations}\label{eq:SCET-RGE}
  \begin{align}
      \frac{\mathrm{d}\tilde{\mathcal {S}}^{[n]}(\nu ,\mu)}{\mathrm{d}\ln\mu}=&\biggr[ -2\Gamma_{\mathrm{cusp}}^{S}(\alpha_s) \ln \frac{ \bar{\nu} \mu}{M_S} +\gamma^S(\alpha_s) \biggr] \tilde{\mathcal {S}}^{[n]}(\nu ,\mu), \\
      \frac{\mathrm{d}\tilde{\mathcal {J}}(\nu ,\mu)}{\mathrm{d}\ln\mu}=& \biggr[ -2\Gamma_{\mathrm{cusp}}^{J}(\alpha_s) \ln\frac{\bar{\nu}\mu^2}{M_J^2} + \gamma^J(\alpha_s) \biggr] \tilde{\mathcal {J}}(\nu ,\mu), \\
      \frac{\mathrm{d} \mathcal {H}^{[n]}(\mu_r,\mu)}{\mathrm{d}\ln\mu}=& \biggr[ -2\Gamma_{\mathrm{cusp}}^{H}(\alpha_s) \ln\frac{\mu M_S}{M_J^2} + \gamma^H(\alpha_s) \biggr] \mathcal {H}^{[n]}(\mu_r,\mu),
  \end{align}
\end{subequations}
where anomalous dimensions obey relations $\Gamma_{\mathrm{cusp}}^{S}=-\Gamma_{\mathrm{cusp}}^{J}=\Gamma_{\mathrm{cusp}}^{H}$ and $\gamma^S+\gamma^J+\gamma^H=0$, and they are universal series in $\alpha_s$,
\begin{align}\label{eq:cusp-function}
\Gamma_{\mathrm{cusp}}^{K}(\alpha_s)=& \sum_{m=0} \biggr(\frac{\alpha_s} {4\pi}\biggr)^{m+1} \Gamma^K_m , \nonumber\\
\gamma^K(\alpha_s)  =& \sum_{m=0} \biggr(\frac{\alpha_s} {4\pi}\biggr)^{m+1} \gamma^K_m, \quad K=S,J,H.
\end{align}
Up to NLL accuracy, the needed coefficients are \cite{Fleming:2003gt,Korchemsky:1987wg}
\begin{eqnarray}\label{eq:cusp-coefficient}
 \Gamma^S_0 &=& -\Gamma^J_0=\Gamma^H_0 = 4C_A ,
 \nonumber\\
\Gamma^S_1 &=& -\Gamma^J_1=\Gamma^H_1 = 4C_A \biggr[ \biggr( \frac{67}{9} - \frac{\pi^2}{3}\biggr)C_A- \frac{20}{9} T_f n_f \biggr], \nonumber\\
\gamma^S_0  &=& 4C_A,
\nonumber\\
\gamma^J_0  &=& 2\beta_0,
\nonumber\\
\gamma^H_0  &=& -4C_A - 2\beta_0.
\end{eqnarray}

 We choose characteristic scales for $\mathcal {H}^{[n]}$, $\mathcal {S}^{[n]}$ and $\mathcal {J}$ as follow
\begin{subequations}\label{eq:scales}
	\begin{align}
	\mu_H&=\mu_r, \\
	\mu_S&=\frac{\mu_r}{\nu}, \\
	\mu_J&=\sqrt{\mu_H\mu_S}=\frac{\mu_r}{\sqrt{\nu}},
	\end{align}
\end{subequations}
then by solving the RGEs in eq. \eqref{eq:SCET-RGE}, we can evolve these functions  from their characteristic scales to the scale $\mu_r$.
Therefore, we get the resumed result
\begin{align}
  \tilde{\sigma}[n](\nu)\vert_{\mathrm{resummed}} &=
  (1+r^\prime)\sigma_{[n]}^{\prime 0}
  \mathcal {H}^{[n]}(\mu_r,\mu_r)
  \tilde {\mathcal {S}}^{[n]}(\nu,\mu_r/\nu)
  \tilde {\mathcal {J}}(\nu,\mu_r/\sqrt{\nu}) \exp \big[h(\mu_r,\chi_1,\chi_2) \big],
\end{align}
where
\begin{align}
  \mathcal {H}^{[n]}(\mu_r,\mu_r)  \tilde{\mathcal {S}}^{[n]}(\nu,\mu_r/\nu) \tilde{\mathcal {J}}(\nu,\mu_r/\sqrt{\nu})
  &=
  1 + \frac{\alpha_s } {4\pi}R_{[n]}(2m_c,r^\prime,\mu_r) + \frac{\alpha_s } {4\pi}
  \biggr(
      -2C_A\gamma_E^2
      \nonumber\\&\hspace{-4.5cm}
      -8C_A\gamma_E\ln\frac{ M_J}{M_S}
      +(4C_A+\beta_0)\gamma_E
      -\frac{\pi^2}{3}C_A
  \bigg)+\mathcal{O}(\alpha_s^2),
\end{align}
and the evolution function is given by
\begin{align}\label{eq:evolution-function}
  h(\mu_r,\chi_1,\chi_2)
  =&\frac{ 2\pi \Gamma_0^S }{\beta _0^2 }\frac{1}{\alpha_s(\mu_r)} \biggr(\left(2 \chi _1-1\right) \ln \left(1-2 \chi _1\right) - 2\left(\chi _2-1\right)\ln \left(1-\chi _2\right) \biggr)
  \nonumber\\&\hspace{-2cm}+
      \frac{\Gamma_0^S}{\beta_0}\left[\ln \left(1-2 \chi _1\right) \ln \left(\frac{\mu_r e^{\gamma_E}}{M_S}\right)-\ln \left(1-\chi _2\right) \ln \left(\frac{\mu_r^2 e^{\gamma_E}}{M_J^2}\right)\right]
  \nonumber\\&\hspace{-2cm}
  -\frac{1}{2\beta_0} \biggr(\gamma_0^J\ln \left(1-\chi _2\right) + \gamma_0^S \ln \left(1-2 \chi _1\right) \biggr)-\frac{\beta _1 \Gamma_0^S}{4 \beta _0^3} \biggr(\ln^2\left(1-2 \chi _1\right)-2 \ln^2\left(1-\chi _2\right)\biggr) \nonumber\\&\hspace{-2cm}
  + \frac{\Gamma_0^S}{2\beta_0^2} \biggl( \frac{\Gamma_1^S}{\Gamma_0^S} -  \frac{\beta_1}{\beta_0} \biggr) \biggr(\ln \left(1-2 \chi _1\right)-2 \ln \left(1-\chi _2\right)\biggr),
\end{align}
with
\begin{align}
\chi_1=\chi_2=\frac{\alpha_s(\mu_r) \beta_0}{4\pi}\ln \nu.
\end{align}
Similar to the SGD, to deal with the Landau singularity, we make following replacement in eq. \eqref{eq:evolution-function}
\begin{align}
\chi_1 \rightarrow  \chi_{1\ast}&=\frac{\alpha_s(\mu_r) \beta_0}{4\pi}\ln  \left(\frac{\nu}{1+\nu/\nu_{1\ast}^L}\right), \quad \nu_{1\ast}^L=\nu_1^L/a,\\
\chi_2 \rightarrow \chi_{2\ast}&=\frac{\alpha_s(\mu_r) \beta_0}{4\pi}\ln  \left(\frac{\nu}{1+\nu/\nu_{2\ast}^L}\right),\quad \nu_{2\ast}^L=\nu_2^L/a,
\end{align}
with branch points $\nu_1^L$ and $\nu_2^L$ given by
\begin{align}
\nu_1^L = \exp\biggr(\frac{2\pi}{\beta_0 \alpha_s(\mu_r)} \biggr) ,\quad
\nu_2^L = \exp\biggr(\frac{4\pi}{\beta_0 \alpha_s(\mu_r)} \biggr) .
\end{align}
The resummed cross section is modified as
\begin{align}\label{eq:resummed-laplace}
  \tilde{\sigma}[n](\nu)\vert_{\mathrm{resummed}} =&
(1+r^\prime)\sigma_{[n]}^{\prime 0} \biggr[1 + \frac{\alpha_s } {4\pi}R_{[n]}(2m_c,r^\prime,\mu_r) + \frac{\alpha_s } {4\pi}
  \biggr(
      -2C_A\gamma_E^2
      -8C_A\gamma_E\ln\frac{ M_J}{M_S}
       \nonumber\\ &
      +(4C_A+\beta_0)\gamma_E
      -\frac{\pi^2}{3}C_A
  \bigg) \biggr]
   \exp \big[h(\mu_r,\chi_{1\ast},\chi_{2\ast}) \big]\tilde{\mathcal {S}}^{\mathrm{mod}}_{[n]}(\nu).
\end{align}
On the other hand, from eq. \eqref{eq:endpoint} we can derive the fixed-order expression of $\tilde{\sigma}[n](\nu)$ as
\begin{align}\label{eq:fixed-order}
\tilde{\sigma}[n](\nu) =& (1+r^\prime)\sigma_{[n]}^{\prime 0}\biggr\{ 1 + \frac{\alpha_s } {4\pi}R_{[n]}(2m_c,r^\prime,\mu_r) + \frac{\alpha_s}{4\pi} \biggr[  \biggr( -8C_A \ln\frac{M_J}{M_S} + 4C_A + \beta_0 \biggr)\ln \bar{\nu}
 \nonumber\\
 &- 2C_A \biggr(\ln^2 \bar{\nu} + \frac{\pi^2}{6}\biggr) \biggr] \biggr\}.
\end{align}
Comparing with eq. \eqref{eq:resummed-laplace}, we can find the single and double logarithms of $\nu$ in eq. \eqref{eq:fixed-order} have been resummed. It should be noted that our choice of the characteristic scales in eq. \eqref{eq:scales} is different from that in refs. \cite{Fleming:2003gt,Sun:2018yam}. Their scales choices not only resum the logarithms of $\nu$, but also try to resum logarithms of $\ln(s/m_c^2)$. However, terms of $\ln(s/m_c^2)$ can only be resummed by using the double-parton fragmentation framework \cite{Kang:2011mg,Kang:2014tta,Kang:2014pya,Lee:2020dza}, which is beyond the scope of this paper.

Using the numerical Laplace inverse transformation, we then obtain the resummed cross section in momentum space as
\begin{align}\label{eq:resummed-momentum}
\frac{\mathrm{d}\hat\sigma[n]}{\mathrm{d}z^\prime}\biggr\vert_{\rm{resummed}}
=&\frac{1}{2\pi i}\int_{c-i \infty}^{c + i\infty} \mathrm{d}\nu e^{ (1-z^\prime) \nu} (\tilde{\sigma}[n](\nu)\vert_{\mathrm{resummed}}).
\end{align}
Combining eqs.~\eqref{eq:resummed-momentum} and the SDCs results, we obtain the NLO+NLL results
\begin{align}\label{eq:NLO+NLL}
  \frac{\mathrm{d}\hat\sigma[n]}{\mathrm{d}z}\biggr\vert_{\mathrm{NLO+NLL}}
  =&
  \frac{\mathrm{d}\hat\sigma[n]}{\mathrm{d}z}
  + \frac{P[r^\prime,z^\prime]}{1+r^\prime}\frac{\mathrm{d}\hat\sigma[n]}{\mathrm{d}z^\prime}\biggr\vert_{\mathrm{resummed}}
  - \frac{1}{1+r^\prime}\frac{\mathrm{d}\hat\sigma[n]}{\mathrm{d}  z^\prime}\biggr\vert_{\mathrm{endpoint}},
\end{align}
where the subtraction of the last term is to avoid double counting between the endpoint and the full $z$ range.

As discussed before, in the pure SGF, only logarithms coming from soft gluon emission are resummed by using the RGEs of SGDs. And the evolution kernels for RGEs are known to $\mathcal{O}(\alpha_s)$. For a comparison, here we also give two partly resummed results in NRQCD+SCET framework, which we denote as $\mathcal {S}$-resum cross section and $\mathcal {J}$-resum cross section. In deriving the $\mathcal {S}$-resum cross section, the evolution of jet function is closed by choosing initial scales as
\begin{align}
\mu_H=\mu_r, \quad  \mu_S=\frac{\mu_r}{ \nu},  \quad \mu_J=\mu_r.
\end{align}
Similarly, for $\mathcal {J}$-resum cross section, the evolution of shape function is closed by following initial scales choice
\begin{align}
\mu_H=\mu_r, \quad  \mu_S=\mu_r,  \quad \mu_J=\frac{\mu_r}{\sqrt{\nu}}.
\end{align}
In addition, in these two cases we only include the $\mathcal{O}(\alpha_s)$ terms of the anomalous dimensions in eq. \eqref{eq:cusp-function}. Then these two partly resummed cross sections read
\begin{subequations}\label{eq:semi-resummed-laplace}
\begin{align}
  \tilde{\sigma}[n](\nu)\Big\vert_{\mathcal {S}-\mathrm{resum}} =&
(1+r^\prime)\sigma_{[n]}^{\prime 0} \biggr[1 + \frac{\alpha_s } {4\pi}R_{[n]}(2m_c,r^\prime,\mu_r) + \frac{\alpha_s } {4\pi}
  \biggr(\beta_0 \ln \bar\nu+2C_A\ln^2\bar\nu
\nonumber\\ &
-8C_A\ln\bar\nu\ln\frac{ M_J}{\mu_r}
      -8C_A\gamma_E\ln\frac{ \mu_r}{M_S}-4C_A\gamma_E^2
      +4C_A\gamma_E
      -\frac{\pi^2}{3}C_A
  \bigg) \biggr]
  \nonumber\\ &\times
   \exp \big[h_S(\mu_r,\chi_{1\ast}) \big]
  \tilde{\mathcal {S}}^{\mathrm{mod}}_{[n]}(\nu),\\
  \tilde{\sigma}[n](\nu)\Big\vert_{\mathcal {J}-\mathrm{resum}} =&
(1+r^\prime)\sigma_{[n]}^{\prime 0} \biggr[1 + \frac{\alpha_s } {4\pi}R_{[n]}(2m_c,r^\prime,\mu_r) + \frac{\alpha_s } {4\pi}
  \biggr(4C_A\gamma_E\ln\frac{M_S^2}{M_J^2} +\beta_0\gamma_E
  \nonumber\\ & + 4C_A\ln\bar\nu
-2C_A\ln^2\bar\nu-2C_A\ln^2\nu+ 4C_A\ln\frac{M_S^2}{\mu_r^2e^{\gamma_E}}\ln\nu
      -\frac{\pi^2}{3}C_A
  \bigg) \biggr]
  \nonumber\\ &\times
   \exp \big[h_J(\mu_r,\chi_{2\ast}) \big]
  \tilde{\mathcal {S}}^{\mathrm{mod}}_{[n]}(\nu),
\end{align}
\end{subequations}
with
\begin{subequations}
\begin{align}
h_S(\mu_r,\chi_{1\ast})=&\frac{ 2\pi \Gamma^S_0}{\beta_0^2}   \frac{1}{\alpha_s(\mu_r)} \biggr( -(1-2\chi_{1\ast})\ln (1-2\chi_{1\ast}) - 2\chi_{1\ast} \biggr)
- \frac{ \gamma_0^S}{2\beta_0} \ln (1-2\chi_{1\ast}) \nonumber\\&+ \frac{ \Gamma_0^S}{\beta_0} \ln (1-2\chi_{1\ast}) \ln \frac{  \mu_re^{\gamma_E}}{M_S},\\
h_J(\mu_r,\chi_{2\ast})=&\frac{2\pi \Gamma^S_0}{\beta_0^2}   \frac{1}{\alpha_s(\mu_r)} \biggr( 2(1-\chi_{2\ast})\ln (1-\chi_{2\ast}) + 2\chi_{2\ast} \biggr)
- \frac{ \gamma_0^J}{2\beta_0} \ln (1-\chi_{2\ast}) \nonumber\\& - \frac{ \Gamma_0^S}{\beta_0} \ln (1-\chi_{2\ast}) \ln \frac{  \mu_r^2e^{\gamma_E}}{M_J^2}.
\end{align}
\end{subequations}
And the results in momentum space are
\begin{subequations}\label{eq:semi-resummed-momentum}
\begin{align}
\frac{\mathrm{d}\hat\sigma[n]}{\mathrm{d}z^\prime}\biggr\vert_{\mathcal {S}-\rm{resum}}
=&\frac{1}{2\pi i}\int_{c-i \infty}^{c + i\infty} \mathrm{d}\nu e^{ (1-z^\prime) \nu} (\tilde{\sigma}[n](\nu)\Big\vert_{\mathcal {S}-\mathrm{resum}}),\\
\frac{\mathrm{d}\hat\sigma[n]}{\mathrm{d}z^\prime}\biggr\vert_{\mathcal {J}-\rm{resum}}
=&\frac{1}{2\pi i}\int_{c-i \infty}^{c + i\infty} \mathrm{d}\nu e^{ (1-z^\prime) \nu} (\tilde{\sigma}[n](\nu)\Big\vert_{\mathcal {J}-\mathrm{resum}}).
\end{align}
\end{subequations}
Finally we denote the partly resummed NLO results as
\begin{subequations}\label{eq:NLO+semi-resum}
\begin{align}
  \frac{\mathrm{d}\hat\sigma[n]}{\mathrm{d}z}\biggr\vert_{\mathcal{S}-\mathrm{resum}}
  =&
  \frac{\mathrm{d}\hat\sigma[n]}{\mathrm{d}z}
  + \frac{P[r^\prime,z^\prime]}{1+r^\prime}
  \frac{\mathrm{d}\hat\sigma[n]}{\mathrm{d}z^\prime}
  \biggr\vert_{\mathcal {S}-\mathrm{resum}}
 - \frac{1}{1+r^\prime}\frac{\mathrm{d}\hat\sigma[n]}{\mathrm{d}  z^\prime}\biggr\vert_{\mathrm{endpoint}},\\
 \frac{\mathrm{d}\hat\sigma[n]}{\mathrm{d}z}\biggr\vert_{\mathcal{J}-\mathrm{resum}}
  =&
  \frac{\mathrm{d}\hat\sigma[n]}{\mathrm{d}z}
  + \frac{P[r^\prime,z^\prime]}{1+r^\prime}
  \frac{\mathrm{d}\hat\sigma[n]}{\mathrm{d}z^\prime}
  \biggr\vert_{\mathcal {J}-\mathrm{resum}}
 - \frac{1}{1+r^\prime}\frac{\mathrm{d}\hat\sigma[n]}{\mathrm{d}  z^\prime}\biggr\vert_{\mathrm{endpoint}}.
\end{align}
\end{subequations}

\section{Phenomenology}\label{sec:Phenomenology}
In this section, we present phenomenological analysis based on our calculations in previous sections. The center-of-mass energy is chosen as $\sqrt{s}=10.6\rm{GeV}$ for B factories. The QED coupling constant is set as $\alpha=1/137$. We determine the value of the coupling constant $\alpha_s(\mu_r)$ by adopting the two-loop RGE formula and setting $\Lambda^{(3)}_{\overline{\rm{MS}}}=388\rm{MeV}$, and we set the renormalization scale $\mu_r=\sqrt{s}/2=5.3\rm{GeV}$. We take the $J/\psi$ mass $M_\psi=3.1\rm{GeV}$ and the heavy quark mass $m_c=1.5\rm{GeV}$. For  nonperturbative model $\tilde F^{\mathrm{mod}}[n](\nu)$ and $\tilde{\mathcal {S}}^{\mathrm{mod}}_{[n]}(\nu)$, we adopt the model function used in \cite{Fleming:2003gt,Fleming:2006cd,Bauer:2001rh,Sun:2018yam}, which in Laplace
space are given by
\begin{subequations}\label{eq:sgf model}
\begin{align}
\tilde F^{\mathrm{mod}}[n](\nu)
=&
N^\psi[n]\left(\frac{\nu\bar\Lambda}{A M_\psi}+1\right)^{-A B},\\
\tilde{\mathcal {S}}^{\mathrm{mod}}_{[n]}(\nu)
=&
\left(\frac{\nu\bar\Lambda}{A M_S}+1\right)^{-A B}.
\end{align}
\end{subequations}
Following ref. \cite{Fleming:2003gt}, we assume that the parameters $A$, $B$ and $\bar \Lambda$ in the model functions are the same for both of the S-wave and P-wave contributions. Furthermore, we also assume that these parameters are the same for the models of SGDs and shape functions. We set $A=1$, $B=2$, $\bar \Lambda=0.3\rm{GeV}$ \cite{Fleming:2003gt}, $a=1.1$ and $x_{\mathrm{min}}=0.5$. $N^\psi[n]$ is the normalization of the model of SGD and we set \cite{Ma:2017xno}
\begin{subequations}
\begin{align}
N^\psi[\COaSz]=& \langle  \mathcal{O}^{J/\psi}(\COaSz)  \rangle,\\
N^\psi[\state{{3}}{P}{J,\lambda}{8}]=& \langle  \mathcal{O}^{J/\psi}(\state{{3}}{P}{0}{8})  \rangle.
\end{align}
\end{subequations}
Then after integrating out $x$, both the differential cross sections in SGF approach and in NRQCD approach can be written in the form
\begin{align}\label{eq:d-cross-section}
  \mathrm{d}\sigma(e^+e^- \to J/\psi + X)
  =
  \mathrm{d}\hat{\sigma}_S
  \langle  \mathcal{O}^{J/\psi}(\COaSz)  \rangle +   \mathrm{d}\hat{\sigma}_P
  \langle  \mathcal{O}^{J/\psi}(\COcPz)  \rangle.
\end{align}
And for the integrated cross
sections we have
\begin{align}\label{eq:total-cross-section}
  \sigma(e^+e^- \to J/\psi + X)
  =
  \hat{\sigma}_S
  \langle  \mathcal{O}^{J/\psi}(\COaSz)  \rangle +   \hat{\sigma}_P
  \langle  \mathcal{O}^{J/\psi}(\COcPz)  \rangle.
\end{align}

\begin{figure}[htb!]
 \begin{center}
 \vspace*{0.8cm}
 \hspace*{-5mm}
 \includegraphics[width=0.45\textwidth]{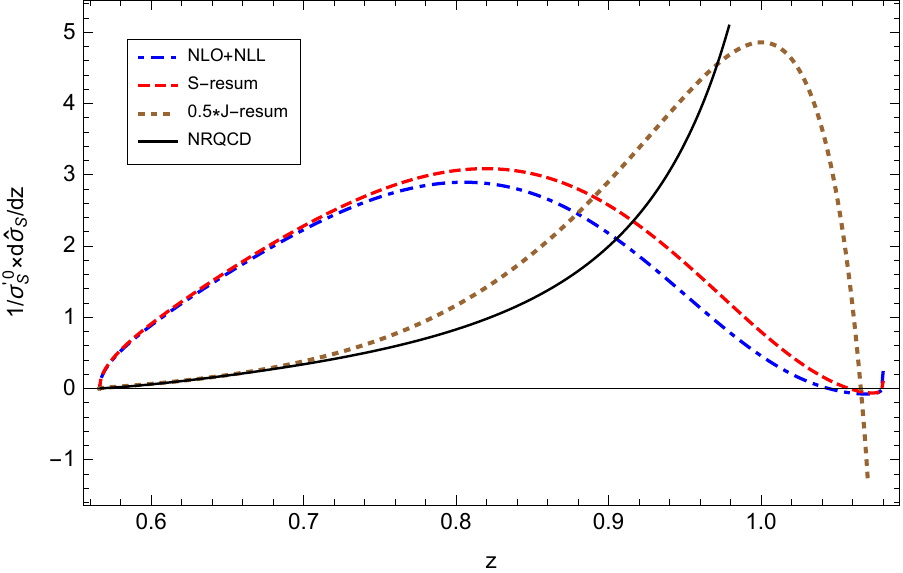}
 \hspace*{5mm}
 \includegraphics[width=0.45\textwidth]{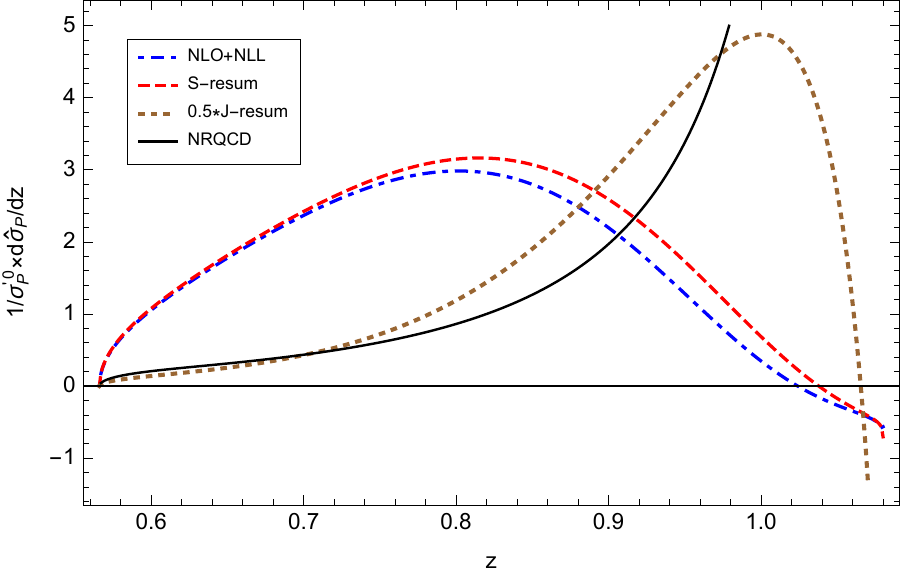}
 \end{center}
 \vspace*{-.5cm}
 \caption{The differential cross section in NRQCD factorization approach with different resummation methods. Left figure is the S-wave contribution and right figure is the P-wave contribution. The $\mathcal {J}$-resum cross section is multiplied by a factor $0.5$. \label{fig:scetfig}}
 \vspace*{0.cm}
\end{figure}

In figure~\ref{fig:scetfig} we show the coefficients of the differential cross
section in NRQCD factorization approach with different resummation methods. We can see that the resummed differential cross sections are much softer than the reult in NRQCD factorization. Moreover, in the NLO+NLL case the unphysical enhancement near the endpoint is cured by taking the resummation and nonperturbative shape function model into account. { But the existence of the next-to-leading power logarithms, like $\ln(1-z^\prime)$, which are not resummed, still drives the differential cross section divergent in the endpoint region.}  We also find the $\mathcal {S}$-resum cross section is close to the result in NLO+NLL method, while the $\mathcal {J}$-resum cross section is deviate from the NLO+NLL result seriously. This phenomenon indicates that, as a good approximation, it makes sense to neglect the evolution of the jet function, as was done in pure SGF.
%%%%%%%%%%%%%%%%%%%%%%%%%%%%%%%%%%%%%%%%%%%%%%%%%%%%%%%%%%%%%%%%%%%%%%%%%%%%
%%%%%%%%%%%%%%%%%%%%%%%%%%%%%%%%%%%%%%%%%%%%%%%%%%%%%%%%%%%%%%%%%%%%%%%%%%%%

\begin{figure}[htb!]
 \begin{center}
 \vspace*{0.8cm}
 \hspace*{-5mm}
 \includegraphics[width=0.45\textwidth]{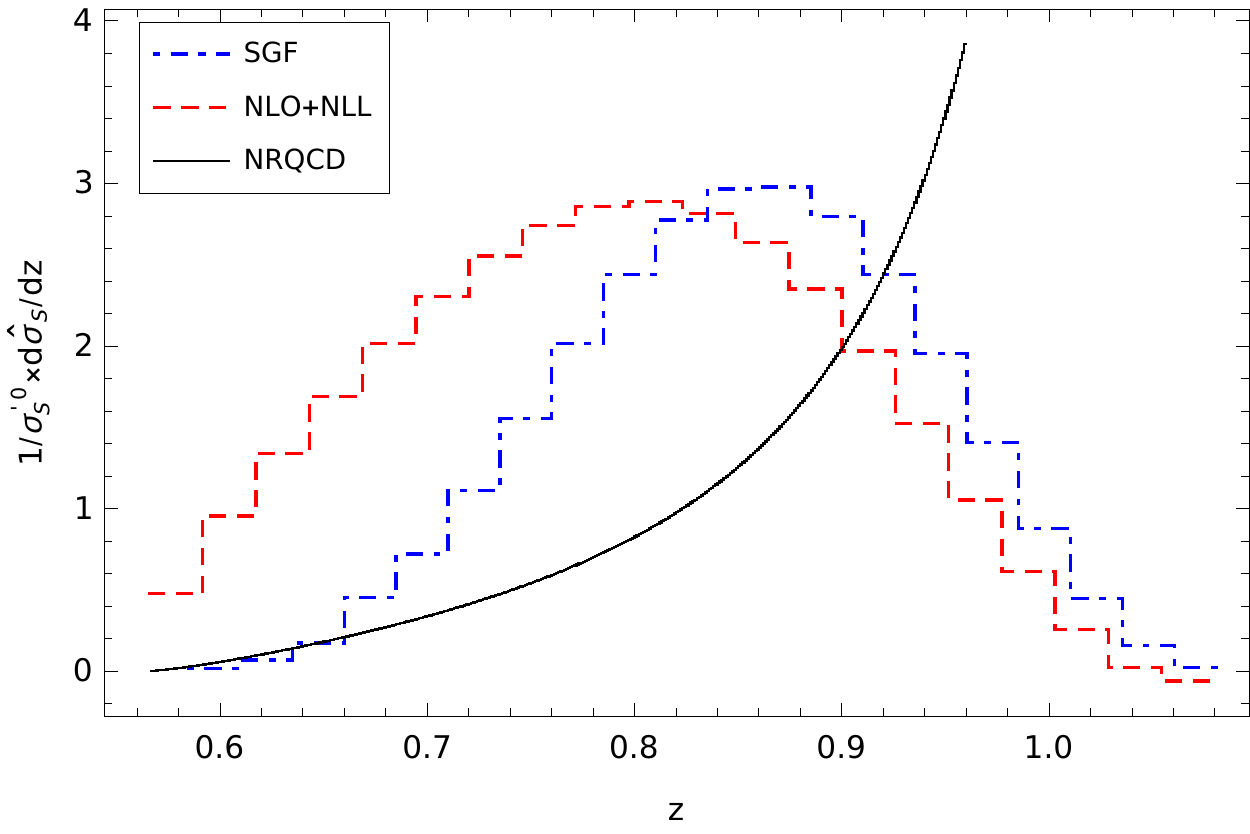}
 \hspace*{5mm}
 \includegraphics[width=0.45\textwidth]{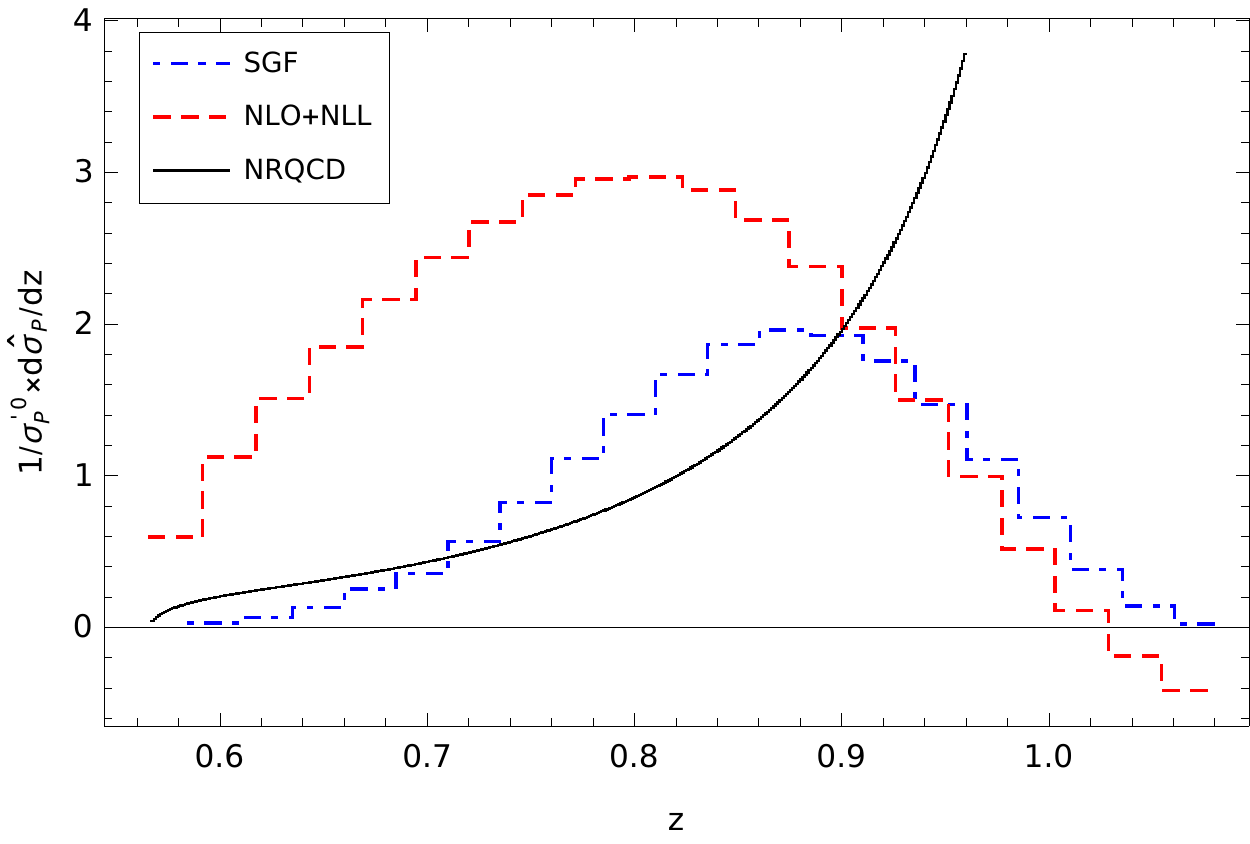}
 \end{center}
 \vspace*{-.5cm}
 \caption{The differential cross
sections in SGF and NRQCD factorization approaches. \label{fig:SGF}}
 \vspace*{0.cm}
\end{figure}

In figure~\ref{fig:SGF} we show the differential cross
sections in SGF comparing with that in NRQCD factorization. Due to the introduction of $x_{\mathrm{min}}$, the $z$ distribution in SGF approach is piecewise function. Therefore, here we calculate the $z$ distributions in $20$ bins, and plot the average value for each bin. We find the peak in NRQCD+NLL results is on the left of that in SGF. We also find that, comparing to the NLO+NLL results, the SGF results are significantly suppressed at moderate $z$, especially for the P-wave case. There are many origins for these differences. First, in solving the RGEs of SGDs, we have included the effects at higher order in $1-x$, which will shift the peak to the right. While in the NRQCD+NLL method these effects have been ignored. Second, as shown in eqs.\eqref{eq:LO-hard-part}, \eqref{eq:NLO-hardpart-S} and \eqref{eq:NLO-hardpart-P}, in SGF approach the hard parts in S-wave and P-wave are proportional
to an overall factor $x$ and $x^3$ (due to the factor $1/M_\psi$ and $1/M_\psi^3$) respectively, which originated from velocity corrections resummation. Such factors suppress the differential cross sections at moderate $z$, especially for the P-wave.
This implies that, even though large logarithms are resummed in NRQCD+SCET approach, there are still significant velocity corrections. Resumming these velocity corrections is the main purpose of the SGF.  Finally, the SGF results included partial contributions at order $\mathcal {O}(\alpha_s^3)$, which come from the convolution of NLO short-distance hard parts
with the resummed SGDs.
%%%%%%%%%%%%%%%%%%%%%%%%%%%%%%%%%%%%%%%%%%%%%%%%%%%%%%%%%%%%%%%%%%%%%%%%%%%%
%%%%%%%%%%%%%%%%%%%%%%%%%%%%%%%%%%%%%%%%%%%%%%%%%%%%%%%%%%%%%%%%%%%%%%%%%%%%

In Table.\ref{tab:coefficient} we list the coefficients of the integrated cross
section in different methods.  We find the $\mathcal {S}$-resum reults are close to the NLO+NLL reults. Besides, both the NLO+NLL results and SGF results are much smaller than the NRQCD reults due to resummation effects. Comparing to the NLO+NLL results, the SGF results are further suppressed.

%%%%%%%%%%%%%%%%%%%%%%%%%%%%%%%%%%%%%%%%%%%%%%%%%%%%%%%%%%%%%%%%%%%%%%%%%%%%
\begin{table}[H]
	\centering
	\begin{tabular}{lcccc}
		\hline \hline
		&NRQCD &NLO+NLL &$\mathcal {S}$-resum &SGF \\
		\hline
		$\hat\sigma_S^{\rm{NLO}}$ ($\rm{pb/GeV^3}$)   &17.677 &7.411 &8.246  &5.986  \\
		\hline
		$\hat\sigma_P^{\rm{NLO}}$ ($\rm{pb/GeV^5}$)   &30.370 &12.691 &14.096  &6.541 \\
		\hline
		\hline
	\end{tabular}
	\caption{\label{tab:coefficient} The coefficients of the integrated cross
section in different methods. }
\end{table}

\begin{figure}[htb!]
 \begin{center}
 \vspace*{0.8cm}
 \hspace*{-5mm}
 \includegraphics[width=0.45\textwidth]{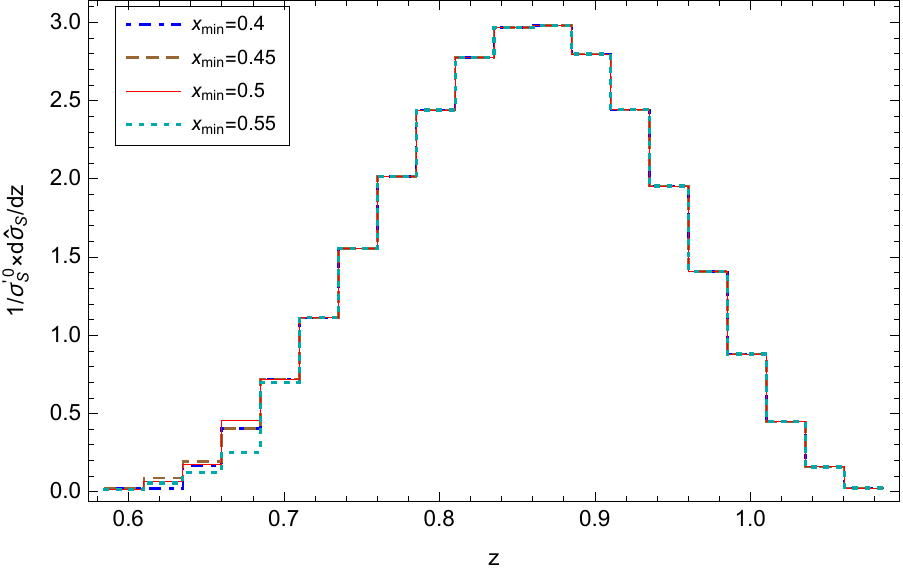}
 \hspace*{5mm}
 \includegraphics[width=0.45\textwidth]{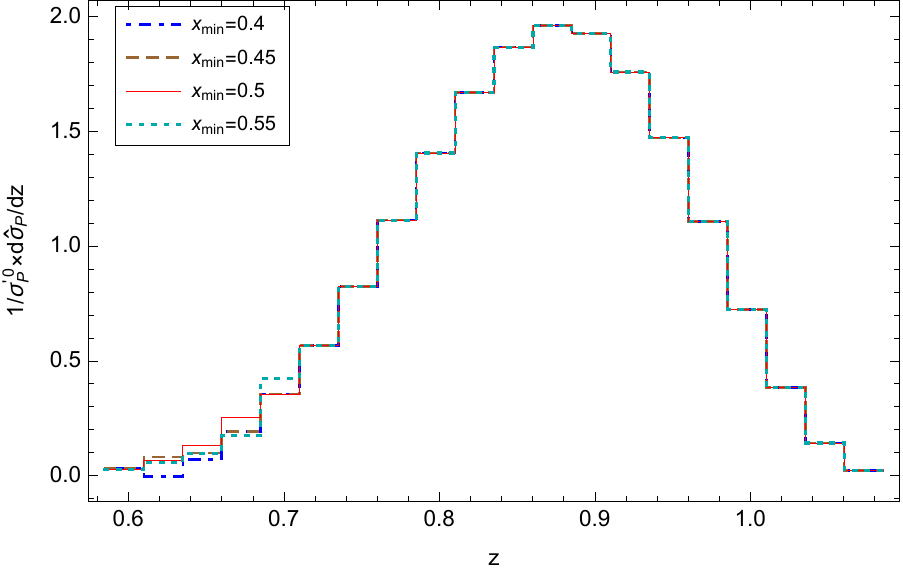}
 \end{center}
 \vspace*{-.5cm}
 \caption{The differential cross
sections in SGF with different $x_{\mathrm{min}}$. \label{fig:xmin}}
 \vspace*{0.cm}
\end{figure}

{In figure~\ref{fig:xmin} we show the $x_{\mathrm{min}}$ dependence of the $z$ distributions by varying $x_{\mathrm{min}}$ from $0.4$
to $0.55$.} We find different choices of $x_{\mathrm{min}}$ can only slightly affect the distribution at moderate $z$ as far as $x_{\mathrm{min}}$  is not too large, e.g., $x_{\mathrm{min}}<0.55$. In Table.\ref{tab:xmin},  $x_{\mathrm{min}}$ dependence of coefficients of cross section is shown. We find differences  are smaller than $2\%$, which confirms our argument in appendix~\ref{ap:cutoff-dependence} that the cross section $\sigma_{J/\psi}$ in SGF is not sensitive to the choice of $x_{\mathrm{min}}$.
%%%%%%%%%%%%%%%%%%%%%%%%%%%%%%%%%%%%%%%%%%%%%%%%%%%%%%%%%%%%%%%%%%%%%%%%%%%%
%%%%%%%%%%%%%%%%%%%%%%%%%%%%%%%%%%%%%%%%%%%%%%%%%%%%%%%%%%%%%%%%%%%%%%%%%%%%
%%%%%%%%%%%%%%%%%%%%%%%%%%%%%%%%%%%%%%%%%%%%%%%%%%%%%%%%%%%%%%%%%%%%%%%%%%%%
\begin{table}[H]
	\centering
	\begin{tabular}{lcccc}
		\hline \hline
		&$x_{\mathrm{min}}$=0.4~~ &$x_{\mathrm{min}}$=0.45~~  &$x_{\mathrm{min}}$=0.5~~ &$x_{\mathrm{min}}$=0.55 \\
		\hline
		$\hat\sigma_S^{\rm{NLO}}$ ($\rm{pb/GeV^3}$)   &5.963 &5.984 &5.986 &5.922   \\
		\hline
		$\hat\sigma_P^{\rm{NLO}}$ ($\rm{pb/GeV^5}$)   &6.471 &6.512 &6.541 &6.521  \\
		\hline
		\hline
	\end{tabular}
	\caption{\label{tab:xmin} The coefficients of the integrated cross
section in SGF with different $x_{\mathrm{min}}$. }
\end{table}
%%%%%%%%%%%%%%%%%%%%%%%%%%%%%%%%%%%%%%%%%%%%%%%%%%%%%%%%%%%%%%%%%%%%%%%%%%%%

Following ref. \cite{Zhang:2009ym}, we define a linear combination of CO LDMEs as:
\begin{align}
  M_k^{X} =
  \langle  \mathcal{O}^{J/\psi}(\COaSz)  \rangle
  +
  k\frac{\langle  \mathcal{O}^{J/\psi}(\COcPz)  \rangle}{m_c^2},
\end{align}
where ``$X$'' denotes factorization approach. If we set the CS contribution to be zero, and
use the CO contribution to saturate the observed
production cross section \cite{Belle:2009bxr}
\begin{align}
	\sigma[e^+e^-\to J/\psi + X_{\mathrm{non}-c\bar c} ]
	=
	0.43\pm0.13\mathrm{pb},
\end{align}
we can get an upper bound for the CO matrix element in each approach. Using the coefficients in Table.\ref{tab:coefficient}, we then obtain
\begin{subequations}\label{eq:CO-ME}
  \begin{align}
    M_{3.9}^{\mathrm{NRQCD}} <& (2.4\pm0.7)\times10^{-2}\mathrm{GeV}^3,\\
    M_{3.9}^{\mathrm{NLO+NLL}} <& (5.8\pm1.8)\times10^{-2}\mathrm{GeV}^3,\\
    M_{2.5}^{\mathrm{SGF}} <& (7.2\pm2.2)\times10^{-2}\mathrm{GeV}^3.
  \end{align}
\end{subequations}
On the other hand, the value of CO LDME $M_k$ extracted from hadron colliders reads \cite{Ma:2010yw}
\begin{align}
    M_{3.9}^{\mathrm{NRQCD,pp}} =& (7.4\pm1.9)\times10^{-2}\mathrm{GeV}^3,
\end{align}
which is about $3$ times larger than the upper bound of $M_{3.9}^{\mathrm{NRQCD}}$. Such a large discrepancy challenges the universality of NRQCD LDMEs. The SGF approach significantly reduces this discrepancy, which provides a hope to solve the universality problem. To this end, we must also describe $J/\psi$ production in hadron colliders using SGF, which will be studied in future works.

\section{Summary}\label{sec:summary}

In summary, in this paper we studied the $J/\psi$ production via color-octet channel, $e^+e^-\to J/\psi(\COcPj,\COaSz) + X_{\mathrm{non}-c\bar c}$, in SGF approach, in which a series of important velocity corrections can be resummed naturally. The corresponding $J/\psi$ energy spectrum is expressed as
a convolution of perturbatively calculable short-distance hard parts with one-dimensional color-octet SGDs, as shown in eq.~\eqref{eq:SGF-form}. We introduced a cutoff $x_{\mathrm{min}}$ for the longitudinal momentum fraction of the
emitted soft gluons to prevent the gluons to be hard. We demonstrated both analytically and numerically that the $x_{\mathrm{min}}$ dependence  is suppressed and can be ignored in the sense of perturbation theory.

We calculated short-distance hard parts analytically up to NLO in $\alpha_s$. We derived and solved RGEs of SGDs, which resums Sudakov logarithms originated from soft gluons emission. Then by adopting a simple model for SGDs at an initial scale, our results are well-behaved near the kinematic endpoint, and they have the same shape as
experimental data for the energy spectrum.
By ignoring color-singlet contribution and using color-octet contributions to saturate the observed production cross section $\sigma[e^+e^-\to J/\psi + X_{\mathrm{non}-c\bar c} ]$, we get a upper bound for the color-octet matrix element $M_k^{X}$, which seems to be consistent with the value extracted from hadron colliders.

However, there are two questions needing to be understood before we can convince ourselves that SGF has provided a reasonable description of $J/\psi$ production. The first question is how important of the  Sudakov logarithms originated from jet functions which have not been resummed in SGF. If this effect is very important, we need to do further job to resum them. The second question is how important of the  velocity-correction terms that have been resummed in SGF. If this effect is not important, we do not really need the SGF approach, but simply using the NRQCD factorization combining with resummation methods to deal with Sudakov logarithms.

To understand the above two questions, we also calculated the same quantity in NRQCD factorization and use NRQCD+SCET approach to resum all encountered Sudakov logarithms. It was found that the full NRQCD+SCET result is very close to the result in which Sudakov logarithms originated from jet functions are not resummed. But if Sudakov logarithms originated from soft gluons emission are not resummed, one will get a result with large deviation from the full result. This answered the first question that  Sudakov logarithms originated from jet functions are not that important. By comparing the  NRQCD+SCET result with SGF result, we still find large difference, which implies that the resummation of velocity-correction terms is significant for phenomenological study.

With the understanding of the above two questions, we conclude that we have provided by far the best theoretical framework to describe $J/\psi$ production in $e^+e^-$ collisions. To further understand the production mechanism of quarkonium, it will be very useful to apply SGF for other processes, like photo- or hadroproduction, in the future.

\section{Acknowledgments}

We thank Kuang-Ta Chao and Xiaohui Liu for many useful discussion. The work is supported in part by the National Natural Science Foundation of China (Grants No. 11875071, No. 11975029), the National Key Research and Development Program of China under Contracts No. 2020YFA0406400.

\begin{appendices}
    \renewcommand{\theequation}{\thesection.\arabic{equation}}
    \setcounter{equation}{0}

    \section{Analytic expressions}
    \label{ap:expression}

    In this appendix, we provide analytic expressions of the functions $\mathrm{d}\sigma^{LO}_{P}/\mathrm{d}z$, $R_S$, $R_P$, $\mathcal {R}_S$, $\mathcal {R}_P$, $\mathcal {G}_S$ and $\mathcal {G}_P$ in section \ref{sec:cross-section} and section \ref{sec:NRQCD}. The perturbative differential cross section in P-wave channel at LO is given by
    {\small
    \begin{align} \label{eq:LO-P}
      &\frac{\mathrm{d}\sigma_P^{LO}}{\mathrm{d}z} ({z},M_{\psi},s,m_c)\nonumber\\
      =&
      \frac{128 \pi ^2 \alpha ^2 \alpha _s e_c^2 }{27 s^4 M_{\psi}^3 (1-r) (1+r) (\frac{M_\psi}{2}-m_c)^2 (\frac{M_\psi}{2}+m_c)^4}\delta(1-\bar z)
      \biggl[\frac{M_{\psi}^{10}}{8}+\frac{3}{4} m_c M_{\psi}^9\nonumber\\
      &+M_{\psi}^8\left(\frac{1}{16} \left(18 \mathcal{T} ^2-60 \mathcal{T} +65\right) m_c^2+\frac{5s}{32}\right)+
      M_{\psi}^7 \left(\frac{3}{16} s (7-2 \mathcal{T} ) m_c+\frac{3}{2} \left(3\mathcal{T} ^2-11 \mathcal{T} +7\right) m_c^3\right)\nonumber\\
      &+
      M_{\psi}^6 \left(-\frac{1}{4} s \left(9 \mathcal{T} ^2-12 \mathcal{T} +2\right) m_c^2+\left(9 \mathcal{T}^2-33 \mathcal{T} +\frac{79}{8}\right) m_c^4+\frac{3 s^2}{16}\right)\nonumber\\
      &+
      M_{\psi}^5\left(\frac{3}{16} s^2 (2 \mathcal{T} +1) m_c-\frac{3}{2} s \left(3 \mathcal{T} ^2-3 \mathcal{T} +5\right)m_c^3+6 \mathcal{T}  (3 \mathcal{T} -7) m_c^5\right)\nonumber\\
      &+
      M_{\psi}^4 \left(\frac{3}{16} s^2 \left(6
      \mathcal{T} ^2-20 \mathcal{T} +9\right) m_c^2+\frac{1}{2} s \left(9 \mathcal{T} ^2-36 \mathcal{T} +8\right)
      m_c^4+3 \mathcal{T}  (15 \mathcal{T} -11) m_c^6\right)\nonumber\\
      &+M_{\psi}^3 \left(\frac{3}{2} s^2 (1-4 \mathcal{T}) m_c^3+3 s \mathcal{T}  (3 \mathcal{T} +2) m_c^5+72 \mathcal{T} ^2 m_c^7\right)\nonumber\\
      &+
      M_{\psi}^2\left(\frac{3}{2} s^2 \left(3 \mathcal{T} ^2-5 \mathcal{T} +1\right) m_c^4+12 s \mathcal{T}  (3 \mathcal{T} -1)
      m_c^6+54 \mathcal{T} ^2 m_c^8\right)+9 s^2 \mathcal{T} ^2 m_c^5 M_{\psi}
      \nonumber\\& +9 s^2 \mathcal{T} ^2m_c^6
      \biggr].
    \end{align}
    }
    The expressions of $R_S$, $R_P$, $\mathcal {R}_S$ and $\mathcal {R}_P$ are given as
    {\small
    \begin{align}
      &R_S(M_\psi, r, \mu_r) \nonumber\\
      =& \frac{1}{4(1 - r) (2 - r)^2 N_c}
      \biggl[
        \frac{4}{3} (r-1) \left(r^2 N_c \left(N_c+2 n_f\right)-8 r \left(N_c
         n_f-N_c^2+3\right)+8 N_c n_f-8 N_c^2+36\right)
         \nonumber\\
         &\times \ln r +\frac{4}{3} (1-r) (r-2)^2 N_c \left(2 n_f-23 N_c\right)\ln
         (1+r)+16 \left(2 r^2 N_c^2-r \left(7N_c^2+2\right)+7 N_c^2+3\right)
         \nonumber\\
         &\times
         (1-r) \ln (1-r)-8 (1-r) (r-2)^2 N_c^2 \big(2\ln r \ln (1+r)-4\ln r\ln (1-r) -4\ln (1-r) \ln (1+r)
         \nonumber\\
         &
         +\ln^2 (1+r)+4\ln^2 (1-r)\big)
         +4 (r-2)^2 \left((r-2) N_c^2+1\right)\biggr(2\ln (2-r) \ln r-\ln^2 (2-r)
         \nonumber\\
         &
         -2\text{Li}_2\left(\frac{r}{2-r}\right)\biggr)+8(r-2)^2 \left(r \left(2N_c^2-1\right)-2\right)\ln \left(1-\sqrt{1-r}\right) \ln \left(1+\sqrt{1-r}\right)
         \nonumber\\
         &
          +2 r (r-2)^2 \ln^2r+24 \sqrt{1-r} (r-2)^2 \ln
         \frac{1-\sqrt{1-r}}{1+\sqrt{1-r}}+\frac{4}{9}
         \biggr(
          3N_c  (r-1) (r-2)^2
          \nonumber\\
          &\times \left(2 n_f-11 N_c\right) \ln \frac{4\mu_r^2}{M_\psi^2}+(1-r) \Big(r^2 \big[2N_c n_f (6 \ln2-5) +N_c^2(31-66 \ln2)
          +90\big]
          \nonumber\\
          &-2 r \big[4N_c n_f (6 \ln2-5)+N_c^2(71-150 \ln2) +9 (19+4 \ln2)\big]
          +4 \big[2N_c n_f (6 \ln2-5)
          \nonumber\\
          & +N_c^2(40-75 \ln2) +27 (3+\ln2)\big]\Big)-3 \pi ^2 (r-2)^2 \big[(5 r-4) N_c^2-1 \big]
         \biggr)
      \biggr],\\
      &R_P(M_\psi, r, \mu_r) \nonumber\\
      =&
      \frac{1}{6 \left(7 r^2+2 r+3\right) N_c}\biggl[
      \biggr(-r^6 N_c \left(31 N_c+14 n_f\right)+r^5 \left(94 N_c n_f+35
      N_c^2+240\right)\nonumber\\
      &+r^4 \left(-230 N_c n_f+353 N_c^2-1248\right)+r^3 \left(250 N_c
      n_f-811 N_c^2+2124\right)\nonumber\\
      &-2 r^2 \left(70 N_c n_f-127 N_c^2+528\right)+4 r \left(22N_c n_f+74 N_c^2-57\right)-48 N_c n_f+48 N_c^2+24\biggr)
      \nonumber\\
      &\times \frac{2 \ln r}{(r-2)^3 (r-1)} +2 \left(7 r^2+2 r+3\right) N_c  \left(2 n_f-23 N_c\right)\ln (1+r)-12 N_c^2 \left(7 r^2+2 r+3\right)
       \nonumber\\
      &\times \biggl( 4\ln^2 (1-r) +  \ln^2(1+r) - 4\ln r \ln (1-r) - 4\ln (1-r) \ln (1+r) + 2\ln r \ln (1+r) \biggr)
      \nonumber\\
      &
      -\frac{12  \left(7 r^3 \left(2
      N_c^2+1\right)+r^2 \left(4 N_c^2-25\right)+r \left(6
      N_c^2-3\right)-3\right)}{(1-r)^{3/2}}\ln \frac{1-\sqrt{1-r}}{1+\sqrt{1-r}}\nonumber\\
      &+\frac{6 \left(3 r^4 N_c^2+r^3 \left(2-8
      N_c^2\right)+r^2 \left(17 N_c^2-11\right)-6 r N_c^2+6 N_c^2-3\right)}{(r-1)^2}\bigg( 2\text{Li}_2\left(\frac{r}{2-r}\right) \nonumber\\
      &+ \ln^2 (2-r) - 2\ln r \ln (2-r) \bigg)-\frac{12}{(r-1)^2}\bigg(r^4 \left(22 N_c^2-3\right)+r^3 \left(7-32N_c^2\right)\nonumber\\
      &+r^2 \left(9-10 N_c^2\right)+r \left(5-4
      N_c^2\right)+6\bigg)\ln \left(1-\sqrt{1-r}\right) \ln
      \left(1+\sqrt{1-r}\right)\nonumber\\
      &+\frac{24  }{(r-2)^3(r-1)}\Big(16 r^6 N_c^2-r^5 \left(93
      N_c^2+20\right)+r^4 \left(191 N_c^2+104\right)-r^3 \left(172 N_c^2+177\right)\nonumber\\
      &+r^2\left(113 N_c^2+88\right)+r \left(19-109 N_c^2\right)+42 N_c^2-2\Big)\ln (1-r)+3 (5-3 r) r \ln^2r\nonumber\\
      &+
       \frac{2}{3 (r-2)^3 (r-1)^2}\biggl(-3 (r-1)^2 (r (7 r+2)+3) (r-2)^3 N_c \left(2 n_f-11 N_c\right) \ln \frac{4\mu_r^2}{M_\psi^2}\nonumber\\
       &+(r-1) \Big(r^6 \big[ N_c^2(217-390 \ln2)
       +14N_c n_f (6 \ln2-5)  +414 \big]\nonumber\\
       &-r^5 \big[N_c^2(1601-3138 \ln2)
       +94 N_c n_f(6 \ln2-5)+2718+720 \ln2 \big]\nonumber\\
       &+r^4 \big[N_c^2(4501-8994 \ln2)+230 N_c n_f(6 \ln2-5)+6498+3744 \ln2 \big]\nonumber\\
       &-r^3 \big[19N_c^2 (329-582 \ln2)+250N_c n_f (6 \ln2-5)
       +6858+6372 \ln2 \big]\nonumber\\
       &+2 r^2 \big[ N_c^2(2561-2796 \ln2)
       +70N_c n_f (6 \ln2-5) +36 (51+44 \ln2) \big] \nonumber\\
       &+4 r \big[  N_c^2(537
       \ln2-791) -22N_c n_f (6 \ln2-5)-576+171 \ln2 \big]\nonumber\\
       &+ 24 \big[
       N_c^2(49-75 \ln2)+2N_c n_f (6 \ln2-5) +54-3 \ln2 \big]\Big)+3
       \pi ^2 (r-2)^3
       \nonumber\\
       &\times
        \big[ N_c^2(39 r^4-64r^3+19r^2-18r+12) - 2 r^3+11
       r^2+3\big]\biggr)
      \biggr],\\
      & \mathcal {R}_S(\bar z,M_\psi, r, \mu_r) \nonumber\\
      =&
      \frac{-2(1+r)}{(r \bar z+\bar z-2)^2}\biggr[ \frac{4 n_f}{3 }
      + \frac{N_c }{3(r \bar z-2 r+{\bar z})^2}\bigg((r+1)^2 (6 r-7) {\bar z}^2-4 \left(9 r^3-4 r^2-10 r+3\right){\bar z}
      \nonumber\\
      &
      +4 \left(9 r^3-10 r^2-3r+3\right)\bigg)
      +\ln \frac{\left(-\sqrt{(r+1)^2 {\bar z}^2-4 r}-r{\bar z}+2 r-{\bar z}\right)^2}{4 r (r+1)(1-\bar z)}
      \nonumber\\
      &
      \times
      \frac{2N_c }{(r {\bar z}-2 r+{\bar z})^3\sqrt{(r+1)^2 {\bar z}^2-4 r}}\bigg((r+1)^5 {\bar z}^5-2 (r+1)^4 (3 r+2) {\bar z}^4+2 (r+1)^3
       \nonumber\\
      &\times
      \left(8 r^2+9 r+3\right)
      {\bar z}^3-2 (r+1)^2 \left(12 r^3+15r^2+12 r+1\right) {\bar z}^2+8 r^2 \left(3 r^3+4 r^2+7 r+6\right) {\bar z}\nonumber\\
      &-4 r \left(3 r^4-2 r^3+6r^2+2 r-1\right)\bigg) \biggr],\\
      & \mathcal {R}_P(\bar z,M_\psi, r, \mu_r)   \nonumber\\
      =&  \frac{4 (1+r) }{(r {\bar z}+{\bar z}-2)^4(7 r^2+2 r+3)}\biggl[
        \frac{2n_f }{3}
        \biggr((r+1)^2 \left(3 r^3-13 r^2-4 r-4\right) {\bar z}^2\nonumber\\
        &+4 \left(-4 r^4+7r^3+10 r^2+2 r+3\right) {\bar z}+2 \left(3 r^4+6 r^3-25 r^2+10 r-6\right)+r (r+1)^3 (r+2){\bar z}^3\biggr)
      \nonumber\\
      &-
      \frac{N_c }{3 (r {\bar z}-2 r+{\bar z})^4}
      \biggr(23 r (r+1)^7 (r+2) {\bar z}^7-(r+1)^6 \left(124 r^3+577 r^2+245 r+20\right) {\bar z}^6
      \nonumber\\
      &
      -2 (r+1)^5\left(78 r^4-1564 r^3-1009 r^2-364 r-39\right) {\bar z}^5
      +2 (r+1)^4 (880 r^5-3032r^4-5533 r^3
      \nonumber\\
      &
      -1110 r^2-823 r-42) {\bar z}^4
      -16 r (r+1)^3 \left(220 r^5-92 r^4-1663r^3-645 r^2-88 r-147\right) {\bar z}^3\nonumber\\
      &+8 r (r+1)^2 \left(333 r^6+1108 r^5-3476 r^4-3248r^3-21 r^2-270 r-222\right) {\bar z}^2\nonumber\\
      &-16 r \left(39 r^8+596 r^7+91 r^6-2332 r^5-1909 r^4-94 r^3-81r^2-102 r-72\right) {\bar z}+32 r (63 r^7
      \nonumber\\
      &
      +84 r^6-362 r^5-34 r^4-39 r^3+39r^2-18 r-9)\biggr)
      - \frac{N_c (1-r) }{(r {\bar z}-2 r+{\bar z})^5 \sqrt{(r+1)^2 {\bar z}^2-4 r} }
      \nonumber\\
      &\times \ln \frac{\left(-\sqrt{(r+1)^2 {\bar z}^2-4 r}-r {\bar z}+2 r-{\bar z}\right)^2}{4 r (r+1)(1-{\bar z})} \biggr(3 (r-1) (r+1)^9{\bar z}^9-2 (r+1)^8 \left(14 r^2-9 r-11\right) {\bar z}^8\nonumber\\
      &+2 (r+1)^7\left(93 r^3-60 r^2-94 r-35\right) {\bar z}^7-2 (r+1)^6 \left(406 r^4-88 r^3-641 r^2-294r-55\right) {\bar z}^6\nonumber\\
      &+8 (r+1)^5 \left(288 r^5+123 r^4-523 r^3-445 r^2-105 r-10\right)
      {\bar z}^5\nonumber\\
      &+16 r^2 (r+1)^3 \left(262 r^5+817 r^4-475 r^3-1091 r^2-617 r-240\right) {\bar z}^3\nonumber\\
      &-4
      (r+1)^4 \left(1018 r^6+1438 r^5-1997 r^4-2485 r^3-1201 r^2-129 r-4\right) {\bar z}^4\nonumber\\
      &-16 r(r+1)^2 \left(137 r^7+932 r^6+12 r^5-1282 r^4-637 r^3-400 r^2-108 r+2\right) {\bar z}^2\nonumber\\
      &+32 r^2 \left(13r^8+258 r^7+457 r^6-234 r^5-647 r^4-318 r^3-181 r^2-90 r-26\right) {\bar z}\nonumber\\
      &-64
      r^2 \left(21 r^7+56 r^6-55 r^5-42 r^4-12 r^3-3 r^2-10 r-3\right) \biggr)\biggr].
    \end{align}
    }
The functions $\mathcal {G}_S$ and $\mathcal {G}_P$ in eq.\eqref{eq:integrated-SDCs} are expressed as
    {\small
    \begin{align}
      &\mathcal {G}_S(2m_c,r^\prime,\mu_r)\nonumber\\
      =&
      \frac{1}{ 36 N_c}
      \biggl[
      \frac{24 \ln (1-r^\prime) \left(r^{\prime2} N_c \left(2 n_f-11
      N_c\right)-2 r^{\prime}\left(4 N_c n_f-25 N_c^2+6\right)+8 N_c n_f-50
      N_c^2+18\right)}{(r^{\prime}-2)^2}\nonumber\\
      &-\frac{24}{(r^{\prime}-2)^2 (r^{\prime}-1)}\Big(r^{\prime 3} N_c \left(2 n_f-11
      N_c\right)+r^{\prime 2} \left(-7 N_c n_f+55 N_c^2-12\right)+r^{\prime } \left(4 N_c n_f-76N_c^2+30\right)\nonumber\\
      &+4 N_c n_f+26 N_c^2-18\Big)\ln r^{\prime}
      -\frac{24 \left(2 N_c n_f-5N_c^2-9\right)+108 r^{\prime} \left(N_c^2+2\right)}{(1-r^{\prime})^{3/2}} \ln\frac{1-\sqrt{1- r^{\prime}}}{1+\sqrt{1- r^{\prime}}} \nonumber\\
      &+144 N_c^2\text{Li}_2(r^{\prime})+\frac{72  \left((r^{\prime}-2)
      N_c^2+1\right)}{r^{\prime}-1} \text{Li}_2\left(\frac{r^{\prime }}{2-r^{\prime }}\right) +\frac{36 \ln^2 (2-r^{\prime}) \left((r^{\prime}-2) N_c^2+1\right)}{r^{\prime}-1}\nonumber\\
      &-\frac{72\ln r^{\prime} \ln (2-r^{\prime}) \left((r^{\prime}-2) N_c^2+1\right)}{r^{\prime}-1}+144 N_c^2 \ln (1-r^{\prime}) \ln r^{\prime}+\frac{9 \ln^2 r^{\prime} \left(r^{\prime} \left(3 N_c^2-2\right)-2 N_c^2\right)}{r^{\prime}-1}\nonumber\\
      &-\frac{36\ln \left(1-\sqrt{1-r^{\prime}}\right) \ln \left(1+ \sqrt{1-r^{\prime}}\right) \left(r^{\prime} \left(7
      N_c^2-2\right)-2 \left(N_c^2+2\right)\right)}{r^{\prime}-1}+\frac{4}{(r^{\prime}-2)^2 (r^{\prime}-1)}
      \nonumber\\
      &\times
      \biggr(-3\left(2n_f-11 N_c\right) (r^{\prime}-2)^2 (r^{\prime}-1) N_c \ln\frac{\mu_r ^2}{m_c^2} -16N_c n_f r^{\prime 3} +12N_c n_f r^{\prime 3} \ln2 \nonumber\\
      &+104N_c n_f r^{\prime 2} -60N_c n_f r^{\prime 2} \ln2 -224N_c n_f r^{\prime} +96N_c n_f r^{\prime} \ln2+160 N_c n_f-48N_c n_f \ln2 \nonumber\\
      &+199N_c^2r^{\prime 3} -66N_c^2 r^{\prime 3} \ln2 -1019N_c^2 r^{\prime 2} +366N_c^2 r^{\prime 2} \ln2 +1670N_c^2 r^{\prime} +3\pi ^2 (r^{\prime}-2)^2 \left(r^{\prime } N_c^2-1\right)\nonumber\\
      &-600N_c^2 r^{\prime} \ln2 -856 N_c^2+300N_c^2 \ln2 +90 r^{\prime 3}-432 r^{\prime 2}-72 r^{\prime 2} \ln2+666 r^{\prime}+180 r^{\prime} \ln2\nonumber\\
      &-324-108 \ln2\biggr)
      \biggr],\\
      &\mathcal {G}_P(2m_c,r^\prime,\mu_r)\nonumber\\
      =&
      \frac{1}{36\left(7 r^{\prime 2}+2 r^{\prime}+3\right) N_c}
      \biggl[
      \frac{36  }{(r^{\prime}-1)^2}\big(3 N_c^2 r^{\prime 4}+\left(2-8 N_c^2\right) r^{\prime 3}+\left(17 N_c^2-11\right) r^{\prime 2}-6 N_c^2 r^{\prime}+6 N_c^2\nonumber\\
      &-3\big)\ln^2 (2-r^{\prime}) -\frac{72 \left(3 N_c^2 r^{\prime 4}+\left(2-8 N_c^2\right) r^{\prime 3}+\left(17 N_c^2-11\right) r^{\prime 2}-6 N_c^2 r^{\prime}+6 N_c^2-3\right) }{(r^{\prime}-1)^2}\ln r^{\prime}
       \nonumber\\
      &
      \times \ln (2-r^{\prime}) +288 \left(3 r^{\prime 2}-2 r^{\prime}+2\right)N_c^2
      \ln (1-r^{\prime}) \ln r^{\prime} +288 \left(3 r^{\prime 2}-2 r^{\prime}+2 \right)N_c^2 \text{Li}_2(r^{\prime}) \nonumber\\
      &+\frac{72  \left(3 N_c^2 r^{\prime 4}+\left(2-8
      N_c^2\right) r^{\prime 3}+\left(17 N_c^2-11\right) r^{\prime 2}-6 N_c^2 r^{\prime }+6 N_c^2-3\right)}{(r^{\prime}-1)^2}\text{Li}_2\left(\frac{r^{\prime}}{2-r^{\prime}}
      \right)\nonumber\\
      &+\frac{9  \left(6 \left(7 N_c^2-1\right) r^{\prime 4}+\left(22-114
       N_c^2\right) r^{\prime 3}+2 \left(51 N_c^2-13\right) r^{\prime 2}+\left(10-19 N_c^2\right) r^{\prime}+4 N_c^2\right)}{(r^{\prime}-1)^2}\ln^2r^{\prime}\nonumber\\
      &-36  \biggl(\left(78 N_c^2-6\right) r^{\prime 4}+\left(14-210 N_c^2\right) r^{\prime 3}+2 \left(89 N_c^2+9\right) r^{\prime 2}+\left(10-91 N_c^2\right) r^{\prime}+12
      \left(N_c^2+1\right)\biggr)\nonumber\\
      &\times\frac{\ln \left(1-\sqrt{1-r^{\prime}}\right) \ln\left(1+\sqrt{1-r^{\prime }}\right)}{(r^{\prime}-1)^2}+\biggl(6 \left(-62 N_c^2+n_f N_c-14\right)r^{\prime 3}+\left(680 N_c^2-68 n_f N_c+300\right)
       \nonumber\\
       &
       \times r^{\prime 2}+\left(-613 N_c^2+100 n_f N_c+36\right) r^{\prime}+74 N_c^2-32 n_f N_c+36\biggr)\frac{6 }{(1-r^{\prime})^{3/2}}\ln\frac{1-\sqrt{1-r^{\prime}}}{1+\sqrt{1-r^{\prime}}}
       \nonumber\\
      &-\biggr(2 \left(8 n_f-65 N_c\right) N_c r^{\prime 7}+\left(1113 N_c^2-98 n_f N_c-240\right) r^{\prime 6}+\left(-3599 N_c^2+150 n_f N_c+1488\right) r^{\prime 5}\nonumber\\
      &+3 \left(1825N_c^2+74 n_f N_c-1124\right) r^{\prime 4}+\left(-3901 N_c^2-942 n_f N_c+3180\right) r^{\prime 3}+2 \big(611 N_c^2+546 n_f N_c\nonumber\\
      &-414\big) r^{\prime 2}-4 \left(122 N_c^2+130 n_f
      N_c+63\right) r^{\prime}+8 \left(40 N_c^2+10 n_f N_c+3\right)\biggr)\frac{12 \ln r}{(r^{\prime}-2)^3 (r^{\prime}-1)^2}\nonumber\\
      &+\biggr(\left(14 n_f-65 N_c\right) N_c r^{\prime 6}+\left(523
      N_c^2-94 n_f N_c-120\right) r^{\prime 5}+\left(-1499 N_c^2+230 n_f N_c+624\right) r^{\prime 4}\nonumber\\
      &+\left(1843 N_c^2-250 n_f N_c-1062\right) r^{\prime 3}+4 \left(-233 N_c^2+35 n_f
      N_c+132\right) r^{\prime 2}+ (358 N_c^2-88 n_f N_c
      \nonumber\\
      &
      +114) r^{\prime} +12 \left(-25 N_c^2+4 n_f N_c-1\right)\biggr)\frac{24 \ln (1-r^{\prime}) }{(r^{\prime}-2)^3 (r^{\prime}-1)}+\biggr(-3 (r^{\prime }-1)^2\left(7 r^{\prime 2}+2 r^{\prime}+3\right)
       \nonumber\\
      &
      \times
      \left(2 n_f-11 N_c\right) N_c (r^{\prime}-2)^3\ln \frac{\mu_r^2}{m_c^2} +3 \pi ^2 (13 N_c^2 r^{\prime 4}-2 \left(4N_c^2+1\right) r^{\prime 3}+\left(11-29 N_c^2\right) r^{\prime 2}
      +14 N_c^2 r^{\prime}
      \nonumber\\
      &
      -2 N_c^2+3) (r^{\prime}-2)^3+(r^{\prime}-1) \Big[2 \left((851-195 \ln2) N_c^2+2n_f N_c (-37+21 \ln2)+207\right) r^{\prime 6}\nonumber\\
      &-2 \left((6743-1569 \ln2) N_c^2+(-581+282 \ln2) n_f N_c+9 (151+40 \ln2)\right) r^{\prime 5}\nonumber\\
      &+\left((42155-8994 \ln2)
      N_c^2+4 (-926+345 \ln2) n_f N_c+3744 \ln2+6498\right) r^{\prime 4}\nonumber\\
      &-\left((66301-11058 \ln2) N_c^2+4 (-1579+375 \ln2) n_f N_c+6372 \ln2+6858\right) r^{\prime 3}\nonumber\\
      &+\left((56126-5592 \ln2) N_c^2+280 (-23+3 \ln2) n_f N_c+72 (51+44 \ln2)\right) r^{\prime 2}\nonumber\\
      &+4 \left((-6457+537 \ln2) N_c^2+4
      (247-33 \ln2) n_f N_c+9 (-64+19 \ln2)\right) r^{\prime }\nonumber\\
      &+8 \left((713-225 \ln2) N_c^2+4 (-35+9 \ln2) n_f N_c-9 (-18+\ln2)\right)\Big]\biggr)\frac{4}{(r^{\prime }-2)^3 (r^{\prime }-1)^2}
      \biggr].
    \end{align}
    }

\section{$x_{\textrm{min}}$ dependence in SGF}
\label{ap:cutoff-dependence}

In this appendix, we discuss the $x_{\textrm{min}}$ dependence of the cross section in SGF approach. To this end, we calculate the quantity $\mathrm{d}\sigma_{J/\psi}/\mathrm{d}x_{\textrm{min}}$.
In following discussion we take the S-wave contribution as an example, while the P-wave contribution can be analyzed similarly. From eq.~\eqref{eq:SGF-form} we can derive the integrated cross section as
 \begin{align}
 \sigma_{J/\psi}
 =& \int \mathrm{d}z \int_{\mathrm{max}[(z+\sqrt{z^2-4r})/2,x_{\mathrm{min}}]}^1 \frac{\mathrm{d}x}{x}   H_{[\COaSz]}(\hat z,M_\psi/x,s,m_c, x_{\mathrm{min}}/x,\mu_f)
 \nonumber\\
 &\times
 F_{[\COaSz] \to \psi }(x,M_{\psi},m_c, \mu_f) + \cdots,
\end{align}
where $\cdots$ denotes the P-wave contribution and we will ignore it . Thus we have
\begin{align}
 \frac{\mathrm{d}\sigma_{J/\psi}}{\mathrm{d}x_{\textrm{min}}}
 =& \int \mathrm{d}z \theta \left(z \leq \frac{x_{\mathrm{min}}^2+r}{x_{\mathrm{min}}} \right) \biggr[  -\frac{1}{x_{\mathrm{min}}}   H_{[\COaSz]}(z/x_{\mathrm{min}},M_\psi/x_{\mathrm{min}},s,m_c,1,\mu_f)
 \nonumber\\
 &\times
 F_{[\COaSz] \to \psi }(x_{\mathrm{min}},M_{\psi},m_c, \mu_f) +  \int_{x_{\mathrm{min}}}^1 \frac{\mathrm{d}x}{x}
    \nonumber\\
 &\times
 \Big(\frac{\mathrm{d}}{\mathrm{d}x_{\textrm{min}}} H_{[\COaSz]}(\hat z,M_\psi/x,s,m_c,x_{\mathrm{min}}/x,\mu_f)\Big)
 F_{[\COaSz] \to \psi }(x,M_{\psi},m_c, \mu_f)\biggr].
\end{align}
According to the matching relation in eq. \eqref{eq:matching-relation} and the perturbative cross section results in section \ref{sec:cross-section}, we have
\begin{align}
H_{[\COaSz]}^{LO}(z,M_\psi,s,m_c,x_{\mathrm{min}},\mu_f)=&\frac{256 \pi ^2 \alpha ^2 \alpha _s  e_c^2 m_c^2 \mathcal{T} ^2  }{3
	s^2 M_{\psi}^3} \frac{1-r}{1+r}\delta(1- \bar{z}),
\end{align}
and
\begin{align}
\frac{\mathrm{d}}{\mathrm{d}x_{\textrm{min}}} H_{[\COaSz]}( z,M_\psi,s,m_c,x_{\mathrm{min}},\mu_f)
=& \frac{1}{x_{\mathrm{min}}} H_{[\COaSz]}^{LO}(z/x_{\mathrm{min}},M_\psi/x_{\mathrm{min}},s,m_c,1,\mu_f)
\nonumber\\
 &\times
 F_{[\COaSz] \to c\bar c[\COaSz] }^{NLO}(x_{\mathrm{min}},M_{\psi},m_c, \mu_f)+ \mathcal {O} (\alpha_s^3).
\end{align}
Then up to order $\alpha_s^2$ we find
\begin{align}\label{eq:xmin-depend}
 \frac{\mathrm{d}\sigma_{J/\psi}}{\mathrm{d}x_{\textrm{min}}}
 =& \int \mathrm{d}z \theta \left(z \leq \frac{x_{\mathrm{min}}^2+r}{x_{\mathrm{min}}} \right) \biggr[  -\frac{1}{x_{\mathrm{min}}}  \biggr( H_{[\COaSz]}^{LO}(z/x_{\mathrm{min}},M_\psi/x_{\mathrm{min}},s,m_c,1,\mu_f)
 \nonumber\\ & +H_{[\COaSz]}^{NLO}(z/x_{\mathrm{min}},M_\psi/x_{\mathrm{min}},s,m_c,1,\mu_f)
 \biggr)
 F_{[\COaSz] \to \psi }(x_{\mathrm{min}},M_{\psi},m_c, \mu_f)
      \nonumber\\
 &+  \int_{x_{\mathrm{min}}}^1 \frac{\mathrm{d}x}{x}
 \frac{x}{x_{\mathrm{min}}} H_{[\COaSz]}^{LO}\Big(z/(xx_{\mathrm{min}}),M_\psi/(xx_{\mathrm{min}}),s,m_c,1/x,\mu_f \Big)
\nonumber\\
 &\times
 F_{[\COaSz] \to c\bar c[\COaSz] }^{NLO}(x_{\mathrm{min}}/x,M_{\psi}/x,m_c, \mu_f)
 F_{[\COaSz] \to \psi }(x,M_{\psi},m_c, \mu_f)\biggr].
\end{align}
Let us first ignore the evolution of $F_{[\COaSz] \to \psi }$, then it is natural to model $F_{[\COaSz] \to \psi }$ as
\begin{align}\label{eq:SGD-mod}
F_{[\COaSz] \to \psi }(x,M_{\psi},m_c, \mu_f)=&\int_{x}^1 \frac{dy}{y} F_{[\COaSz] \to c\bar c[\COaSz]}(x/y,M_\psi/y,m_c, \mu_f) F^{\textrm{mod}}[\COaSz](y )\nonumber\\
&\hspace{-4cm}  =F^{\textrm{mod}}[\COaSz](x ) + \int_{x}^1 \frac{dy}{y} F_{[\COaSz] \to c\bar c[\COaSz]}^{NLO}(x/y,M_\psi/y,m_c, \mu_f) F^{\textrm{mod}}[\COaSz](y ),
\end{align}
where $F^{\textrm{mod}}[\COaSz](y )$ is the inverse Laplace transform of the model $\tilde F^{\mathrm{mod}}[\COaSz](\nu)$ in eq. \eqref{eq:Resummed-SGD},
\begin{align}
F^{\textrm{mod}}[\COaSz](y ) =&    \frac{1}{2\pi i}\int_{c-i \infty}^{c + i\infty} d\nu e^{ (1/y-1) \nu} \tilde F^{\textrm{mod}}[\COaSz](\nu).
\end{align}
We assume $F^{\textrm{mod}}[\COaSz](y )$ to be  peaked
around $y=1$ and vanished at small and moderate $y$, i.e. $F^{\textrm{mod}}[\COaSz](x_{\textrm{min}})\sim 0$. Besides, we assume the model function satisfies the moment relations~\cite{Chen:2021hzo,Fleming:2003gt,Fleming:2006cd}
\begin{align}\label{eq:moments}
\int_0^1 \frac{dy}{y^2}   F^{\textrm{mod}}[\COaSz](y ) =& \mathcal {O} (1), \nonumber\\
\int_0^1 \frac{dy}{y^2}  \biggr(\frac{1}{y}-1 \biggr)   F^{\textrm{mod}}[\COaSz](y ) =& \mathcal {O} (\bar\Lambda/M_\psi) , \nonumber\\
\int_0^1 \frac{dy}{y^2}  \biggr(\frac{1}{y}-1 \biggr)^2  F^{\textrm{mod}}[\COaSz](y) =& \mathcal {O} (\bar\Lambda^2/M_\psi^2),
\end{align}
with $\bar\Lambda\sim \Lambda_{\mathrm{OCD}}$. Inserting eq. \eqref{eq:SGD-mod} into eq. \eqref{eq:xmin-depend} we then obtain
\begin{align}
 \frac{\mathrm{d}\sigma_{J/\psi}}{\mathrm{d}x_{\textrm{min}}}
 =& \int \mathrm{d}z \theta \left(z \leq \frac{x_{\mathrm{min}}^2+r}{x_{\mathrm{min}}} \right) \biggr[  -\frac{1}{x_{\mathrm{min}}}  H_{[\COaSz]}^{LO}(z/x_{\mathrm{min}},M_\psi/x_{\mathrm{min}},s,m_c,1,\mu_f)
 \nonumber\\ & \times \int_{x_{\mathrm{min}}}^1
 \frac{dy}{y} F_{[\COaSz] \to c\bar c[\COaSz]}^{NLO}(x_{\mathrm{min}}/y,M_\psi/y,m_c, \mu_f) F^{\textrm{mod}}[\COaSz](y )
      \nonumber\\
 &+  \int_{x_{\mathrm{min}}}^1 \frac{\mathrm{d}y}{y}
 \frac{y}{x_{\mathrm{min}}} H_{[\COaSz]}^{LO}\Big(z/(yx_{\mathrm{min}}),M_\psi/(yx_{\mathrm{min}}),s,m_c,1/y,\mu_f \Big)
 \nonumber\\
 &\times F_{[\COaSz] \to c\bar c[\COaSz]}^{NLO}(x_{\mathrm{min}}/y,M_\psi/y,m_c, \mu_f)
 F^{\textrm{mod}}[\COaSz](y )\biggr]+ \mathcal {O} (\alpha_s^3).
\end{align}
Expanding the functions $H_{[\COaSz]}^{LO}$ and $F_{[\COaSz] \to c\bar c[\COaSz]}^{NLO}$ in $(1/y-1)$ and using eq.~\eqref{eq:moments}, it is easy to find
\begin{align}
 \frac{\mathrm{d}\sigma_{J/\psi}}{\mathrm{d}x_{\textrm{min}}}
 =& \mathcal {O} (\alpha_s^3)+\mathcal {O} (\alpha_s^2 \bar\Lambda/M_\psi),
\end{align}
which can be ignored in perturbative calculation.
Because the difference between the resummed SGD and the SGD modeled in eq.~\eqref{eq:SGD-mod} is $\mathcal {O} (\alpha_s^3)$, the above equation is also valid for resummed SGD. Thus we argue that the $x_{\textrm{min}}$ dependence of $\sigma_{J/\psi}$ is $\alpha_s$ or $\Lambda_{\mathrm{QCD}}/M_\psi$ suppressed.

\end{appendices}

% references
%\bibliographystyle{utphysMa}
%\bibliography{bibTex1.5}

\providecommand{\href}[2]{#2}\begingroup\raggedright\endgroup

\end{document}